\documentclass[sigconf,nonacm]{acmart}
\usepackage{pifont} %

\usepackage{enumitem}

\usepackage{tikz}
\usepackage{amsmath}
\usepackage{amsthm}
\usepackage{url}

\usepackage{multirow}
\usepackage{graphicx}
\usepackage{subfig}
\usepackage{xcolor}
\usepackage{tabularx}
\newtheorem{theorem}{Theorem}
\newtheorem{lemma}{Lemma}
\newtheorem{definition}[theorem]{Definition}
\newtheorem{example}[theorem]{Example}

\newenvironment{ditemize}{%
\begin{list}{{\bf $\bullet$}}{%
\setlength{\itemsep}{0pt}\setlength{\rightmargin}{0pt}%
\setlength{\leftmargin}{1.2em}\setlength{\parsep}{0pt}}}{
\end{list}}

\newcommand{\reducedstrut}{\vrule width 0pt height .9\ht\strutbox depth .9\dp\strutbox\relax}
\newcommand{\gray}[1]{%
  \begingroup
  \setlength{\fboxsep}{0pt}%
  \colorbox{lightgray}{\reducedstrut#1\/}%
  \endgroup
}

\newcommand{\white}[1]{%
  \begingroup
  \setlength{\fboxsep}{0pt}%
  \colorbox{white}%
  \endgroup
}
\newcommand{\system}{\textsf{TMIA}}
\newcommand{\s}{\mathbf{s}}
\usepackage{booktabs}
\begin{document}
\author{Jie Fu}
\email{jfu13@stevens.edu}
\affiliation{%
  \institution{Stevens Institute of Technology}
  \city{Hoboken}
  \state{New Jersey}
  \country{USA}}

\author{Xiangpeng Deng}
\email{xdeng12@stevens.edu}
\affiliation{%
  \institution{Stevens Institute of Technology}
  \city{Hoboken}
  \state{New Jersey}
  \country{USA}}

\author{Yuan Hong}
\email{yuan.hong@uconn.edu}
\affiliation{%
  \institution{University of Connecticut}
  \city{Storrs}
  \state{Connecticut}
  \country{USA}}

\author{Wendy Hui Wang}
\email{hwang4@stevens.edu}
\affiliation{%
  \institution{Stevens Institute of Technology}
  \city{Hoboken}
  \state{New Jersey}
  \country{USA}}

\renewcommand{\shortauthors}{Fu et al.}
\title[Inferring Temporal Membership in Continually Updated Models]{Timing is Everything: Inferring Temporal Membership in Continually Updated Models}

\begin{abstract}
Continual model updating has become a common practice in the deployment of  machine learning systems. However, the dynamic training behaviors give rise to new risks not sufficiently addressed by existing studies. 
In this paper, we present the first systematic study of temporal privacy risk in continually updated deep learning models.
We formalize the notion of \textit{temporal membership} in continually updated models. 
We then define the \textit{temporal membership inference attack}, which seeks  to infer whether and when individual samples were involved in model training,  at both timestamp-level and interval-level granularities.
To this end, we propose \system, a novel black-box attack that leverages bidirectional LSTMs to infer temporal membership from model outputs observed across successive updates.
Through extensive experiments on three language models and three image classification models, we show that \system\ is highly effective and significantly outperforms existing baselines. Finally, we evaluate three potential defenses, revealing the trade-off between privacy protection and model utility. 
\end{abstract}


\maketitle

\section{Introduction}
Modern deep learning models are rarely static. In practice, deployed models are continually updated to incorporate newly collected data \cite{derakhshan2019continuous,parisi2019continual}, remove outdated or erroneous samples, or satisfy regulatory requirements such as the right to be forgotten \cite{regulation2018general,pardau2018california}. These updates are often performed repeatedly over the lifetime of a model \cite{thota2022mlops}. As a result, a single deployed system typically consists of a sequence of model snapshots evolving over time. 

While continual updates improve model adaptability, they introduce new privacy risks. In particular, \textit{membership inference attacks (MIAs)} \cite{shokri2017membership,yeom2018privacy},  which seek to determine whether specific data samples were used to train a model, will take on new forms in the dynamic setting. Beyond inferring whether a sample was ever used for training, an adversary may attempt to infer when a sample entered or left the training process. Revealing not only membership but also its timing can expose additional information about a model's training dynamics. For example, in a continually updated language model, identifying the time interval during which a sensitive document was included in training may reveal details about data retention or update policies. 

Although several prior works have proposed MIAs against models with updates ~\cite{jagielski2023combine,michel2026sequential,naderloui2025rectifying,hayes2024inexact,chen2021machine}, none addresses the membership privacy risks considered in our setting for the following reasons.
First, these studies assume that membership changes occur in only one direction. Specifically, they consider either insertion-only updates~\cite{jagielski2023combine,michel2026sequential}\footnote{Although \cite{jagielski2023combine,michel2026sequential} do not explicitly state an insertion-only assumption, their problem formulation implicitly assumes that training data are updated by inserting new samples.} or removal-only updates~\cite{naderloui2025rectifying,hayes2024inexact,chen2021machine}. 
Thus they are unsuitable for settings where both insertion and deletion occur during model updates. 
Second, these works assume that each sample undergoes at most one membership transition and thus they cannot capture repeated membership changes, where the same sample may be inserted into and removed from the training set multiple times. 
Third, most existing approaches~\cite{naderloui2025rectifying,hayes2024inexact,chen2021machine,jagielski2023combine} infer membership changes by comparing only two consecutive model snapshots. As shown in Section~\ref{sc:characterization}, such pairwise analysis fails to capture the temporal dependencies across a sequence of model updates, resulting in inaccurate reconstruction of a sample's membership history. In particular, these methods may infer temporally inconsistent update sequences, such as predicting that a sample was removed before it was ever inserted. Table~\ref{tab-comparsion} summarizes the key differences between \system\ and these prior studies.

 \begin{table}[!t]
\centering
\caption{Comparison between \system\ and existing MIAs against models with updates. INS: Insertion; DEL: deletion; $f_i, f_{i+1}$: consecutive model snapshot pair;  $f_{0}, f_1, \dots, f_{N}$: Full sequence of model snapshots.}
\label{tab-comparsion}
\scalebox{0.9}{
\begin{tabular}{l|l|l|l}
\hline
\textbf{Study} 
& \textbf{Operation} 
& \textbf{Frequency} 
& \textbf{Attack Input} \\ \hline

\cite{jagielski2023combine}
& INS only
& One-time
& $f_i, f_{i+1}$ \\ \hline

\cite{naderloui2025rectifying,hayes2024inexact,chen2021machine}
& DEL only
& One-time
& $f_i, f_{i+1}$ \\ \hline

\cite{michel2026sequential}
& INS only
& One-time
& $f_0, f_1, \dots, f_N$ \\ \hline

Ours
& INS \& DEL
& Repeatedly 
& $f_0, f_1, \dots, f_N$ \\ \hline
\end{tabular}
}
\end{table}

In this paper, we present the first systematic study of temporal privacy risks in continually updated deep learning models. We consider a general update setting that accommodates both data insertions and removals, as well as single-update and repeated-update scenarios. We formalize the problem of \textit{temporal membership inference} at two levels of granularity. At the \textit{timestamp level}, the adversary seeks to determine whether a target sample belongs to the training set at a specific point in time. At the \textit{interval level}, the adversary aims to infer the time interval(s) during which the sample participates in training, identifying its \textit{entry and exit points} (i.e., when it is inserted into or removed from the model). 

To address this problem, we propose \system, a novel attack framework that leverages temporal patterns in model outputs across the sequence of model snapshots. We made the following contributions:  
\begin{ditemize}
\item We introduce the notion of \textit{temporal membership} and formally define timestamp-level and interval-level membership inference for continually updated models.

\item We develop \system, a black-box temporal membership inference attack that leverages a bidirectional LSTM to accurately predict a sample’s entry and exit points from its confidence-score trajectories across model updates. Beyond identifying entry and exit points, \system\ constructs \textit{valid membership intervals} that respect the temporal constraints, such as correct ordering and exit-entry matching between these points. 

\item We empirically evaluate \system\ on both language models and deep image classification models with various update algorithms. Our results demonstrate that \system\ is highly effective, with F1-scores at both timestamp-level and interval-level outperforming the baseline approaches significantly. Furthermore, we demonstrate that \system\ remains effective when attacking large generative language models. 

\item We extend three widely-used defense mechanisms for static models, namely,  \textit{label-only outputs}, \textit{dropout}, and \textit{differential privacy}, to the continual update setting, and evaluate their effectiveness  against \system. Our results show that  these defenses can reduce \system's privacy risks, but often at the cost of model utility, highlighting the inherent trade-off between protection and model performance in continuously updated models. 

\end{ditemize}

\section{Background and Related Work}
\subsection{Long Short-Term Memory Network}\label{sec-LSTM}
A Long Short-Term Memory (LSTM) network is a class of gated recurrent neural
networks designed to model sequential data with long-range
dependencies via gated memory cells~\cite{hochreiter1997long}. 
At each time step $t_i$, an LSTM maintains a hidden state $\mathbf{h}_i$ and a cell state $\mathbf{c}_i$, 
which are updated through nonlinear gated transformations of the current input $\mathbf{x}_i$, 
the previous hidden state $\mathbf{h}_{i-1}$, and the previous cell state $\mathbf{c}_{i-1}$.

A \textit{bidirectional LSTM }(BiLSTM) extends the LSTM model by processing a sequence 
in both forward and backward directions~\cite{graves2005framewise}. A BiLSTM model consists of two independent LSTM networks: a \textit{forward LSTM }that processes the sequence left-to-right, and a \textit{backward LSTM }that processes the sequence right-to-left. 
The hidden and cell states at the timestamp $t$ for each LSTM are computed accordingly:

\begin{equation}
\label{eqn:lstm}
\begin{aligned}
(\overrightarrow{\mathbf{h}}_i, \overrightarrow{\mathbf{c}}_i)
&=
\mathrm{LSTM}_{\mathrm{fwd}}
(\mathbf{x}_i, \overrightarrow{\mathbf{h}}_{i-1}, \overrightarrow{\mathbf{c}}_{i-1}), \\
(\overleftarrow{\mathbf{h}}_i, \overleftarrow{\mathbf{c}}_i)
&=
\mathrm{LSTM}_{\mathrm{bwd}}
(\mathbf{x}_i, \overleftarrow{\mathbf{h}}_{i+1}, \overleftarrow{\mathbf{c}}_{i+1}).
\end{aligned}
\end{equation}

Finally, the BiLSTM combines the two hidden states:
\begin{equation}
\mathbf{h}_i
=
\big[\overrightarrow{\mathbf{h}}_i \,;\, \overleftarrow{\mathbf{h}}_i\big].
\end{equation}

\subsection{Privacy Leakage from Model Updates}

{\bf MIAs against Model Updates.}
\textit{Membership inference attack (MIA)} is one of the most common privacy attacks against machine learning models \cite{rigaki2023survey}. It aims to determine whether a specific data instance was included in the training set of a target model~\cite{shokri2017membership,salem2018ml}. 
Jagielski et al.~\cite{jagielski2023combine}  developed the first MIA to attack models with updates. They considered both single-update and multiple-update settings and developed two attacks: (1) {\em Membership inference},  where the adversary aims to infer whether a specific data sample is included in the update set at a particular time point; and (2) \textit{Entry inference}, which aims to infer which time epoch the data sample enters the dataset. 
Michel et al.~\cite{michel2026sequential} 
proposed the \textit{Sequential Membership Inference} (SeMI) attacks that exploit the sequence of models and injecting a target canary at a controlled insertion time. They derived the optimal SeMI attack as a likelihood-ratio (LR) test, and designed a LR-based white-box attack and a loss-based black-box attack. Both studies \cite{jagielski2023combine, michel2026sequential} consider data insertions only. Since insertion-only updates is a special case of the update scenario that we consider, they are not directly applicable to our setting where both insertion and removal operations are allowed. 

{\bf MIAs against Machine Unlearning Models.} In practice, model updates often only involve data deletions. \textit{Machine unlearning} (MU) \cite{bourtoule2021machine} enables efficient removal of the influence of specific samples from a trained model without retraining from scratch~\cite{nguyen2022survey,shaik2024exploring}. 
Recent studies have shown that MU can introduce privacy risks. By exploiting the difference between the original and unlearned models, the adversary can infer whether the target sample belongs to the removed set ~\cite{chen2021machine,naderloui2025rectifying,hayes2024inexact,deepak2025identifying,du2024textual,wu2025unlearned}. 
Since these works focus solely on inferring single-removal events, they are not directly applicable to the more general update scenarios considered in this work, where insertions and removals occur continuously.

\section{Problem Formulation} \label{sec-prob}

In this paper, we focus on deep neural networks for classification as our target models. Given an input sample, the target model outputs a \textit{posterior probability vector} over the predefined set of classes, where each entry represents the model’s predicted probability for the corresponding class. 

\subsection{Temporal Membership}
Consider a classification model $f_0$ initially trained on a dataset $D_0$ by running a learning algorithm $A_{\text{train}}$. 
The model $f_0$ is continuously updated through a set of updates $\Delta_1, \dots, 
\Delta_N$ at timestamps$
\mathcal{T} = \{t_1, t_2, \ldots, t_N\}$, yielding a corresponding sequence of model snapshots $\{f_{1}, f_{2}, \ldots, f_{N}\}$.
Each snapshot $f_i$ is obtained by an update algorithm $A_{\text{update}}$ on $f_{i-1}$ and $\Delta_{i}$: $f_i = A_{\text{update}}(f_{i-1},\Delta_{i}).$
Each \textit{update set }$\Delta_i$ consists of two disjoint sets $\Delta_i^+$ and $\Delta_i^-$, where the \textit{insertion set}  $\Delta_i^+$ includes samples inserted into the model $f_{i-1}$, and the \textit{removal set} $\Delta_i^-$ includes samples whose influence is  removed from $f_{i-1}$. 
Either $\Delta_i^+$ or $\Delta_i^-$ may be empty, corresponding to updates that involve only insertions or removals. 
We also allow samples to be repeatedly inserted or removed across different updates. We define the \emph{latest update time} of a sample $x$ up to timestamp $t_i$, denoted as $\eta^x_i$, as follows: 
$\eta^x_i = \max \big\{\, j \le i \;\big|\; x \in \Delta_j^+ \cup \Delta_j^- \,\big\},$ with the convention that $\eta^x_i$ is undefined if $x$ has never been updated.

\begin{definition}[{Temporal membership}]
We say that $x$ is a \emph{member} of model snapshot $f_i$ if either
(i) $x \in D_0$ and $\eta_i^x$ is undefined, or
(ii) the most recent update involving $x$ up to and including snapshot $f_i$ is an insertion, i.e., $x \in \Delta_{\eta_i^x}^{+}$.
Otherwise, $x$ is a \emph{non-member} of $f_i$.
\end{definition}

Although the update-wise MIA proposed in~\cite{jagielski2023combine} aims to infer whether a sample belongs to a specific update set, whereas our goal is to infer whether a sample is a member of a given model snapshot, these two notions of membership are equivalent, as~\cite{jagielski2023combine} considers only data insertions. 

\textbf{Temporal Membership Interval.}
Given a sequence of observed model snapshots, we define the
\emph{temporal membership intervals} of a sample $x$ as the
time periods during which $x$ is a member of the corresponding
model snapshots. Formally, let
$I=[t_{\mathrm{start}},t_{\mathrm{end}}]$
denote a temporal membership interval of $x$, where
$t_{\mathrm{start}}$ is the first timestamp at which $x$
becomes a member of the model, and $t_{\mathrm{end}}$ is the
last timestamp at which $x$ remains a member. Specifically, if $x$ is inserted through the update set
$\Delta_a^+$, then $t_{\mathrm{start}}=t_a$. If $x$ is
subsequently removed through the update set $\Delta_b^-$,
then $t_b$ is the first timestamp at which $x$ becomes a
non-member. Accordingly, we define $t_{\mathrm{exit}}=t_b$ and $t_{\mathrm{end}}=t_{b-1}$. 
Thus, $t_{\mathrm{exit}}$ is the timestamp immediately after
$t_{\mathrm{end}}$ in the model-snapshot sequence. In the following, we refer to $t_{\mathrm{start}}$ as an
\textbf{entry point} and $t_{\mathrm{exit}}$ as an
\textbf{exit point}. All other timestamps at which the
membership status of $x$ does not change are referred to as
\textbf{non-change points}. If $x$ is a member of the initial model $f_0$, we define
$t_0$ as the timestamp corresponding to $f_0$ and set
$t_{\mathrm{start}}=t_0$. 

Since a sample can enter and exit  the training set multiple times, its temporal membership may cover a set of disjoint intervals $\mathcal{I}^*(x) = \{I^*_1, \dots, I^*_m\}$,
where each interval represents the contiguous range of snapshots in which $x$ is included in the training process. 
For two consecutive intervals
$I_i=[t_a,t_b]$ and $I_{i+1}=[t_c,t_d]$, we require
$c\ge b+2$, ensuring that at least one non-member snapshot
separates them. otherwise, the two intervals are contiguous and are merged into $[t_a,t_d]$.
The set of membership timestamps included in $\mathcal{I}^*(x)$ is given by
$T(x) = \bigcup_{I \in \mathcal{I}^*(x)} T(I),$   
where $T(I)$ denotes the timestamps covered by the interval $I$. 

\begin{figure}[t!]
\centering
  \includegraphics[width=1.0\linewidth]{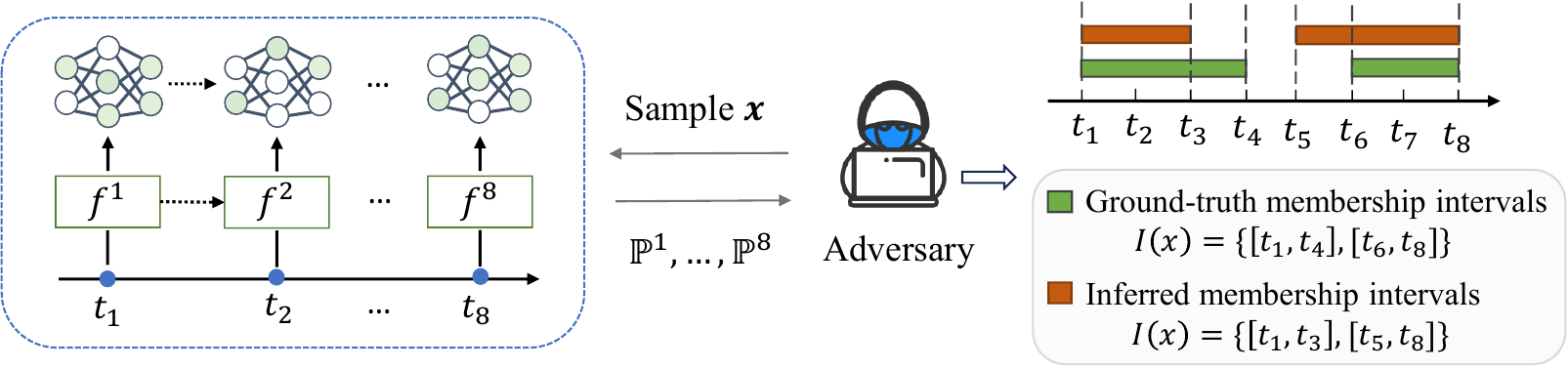}
		\caption{An example of the ground-truth and inferred temporal membership intervals.}
        \label{figure:problem}        	
    \vspace{-0.15in}
\end{figure}

\subsection{Threat Model}

{\bf Attack Goal.} We consider a black-box attacker modeled as an external, untrusted party with legitimate query access to the deployed model, but no involvement in its training. An example of such attacker is a client (user) of a machine learning service (e.g., an API consumer) who can submit inputs and observe outputs. Specifically, given a target sample $x$, the adversary's objective is to infer its \textit{temporal membership intervals} of $x$: 
$\mathcal{I}(x) = \{{I}_1, {I}_2, \ldots, {I}_{k}\}$. 
The number of inferred intervals may not match that of the ground-truth interval(s), allowing the adversary to infer either more or fewer membership intervals than actually present. Figure \ref{figure:problem} shows an example of the ground-truth and inferred temporal membership intervals. 
The inferred membership intervals can be used to characterize privacy risks at different granularity. Specifically, at the {\em timestamp-level}, the adversary infers that the sample $x$ was used by the model at timestamp $t$ if $t$ lies within any interval in $\mathcal{I}(x)$. At the {\em interval-level}, the adversary infers that $x$ was included in the model within a time window $I$ if there is a predicted interval $I_i\in\mathcal{I}(x)$ such that $I\subseteq I_i$. 

Temporal membership inference can leak sensitive information that goes beyond whether a user’s data was ever used. In many real-world deployments, this temporal dimension is tightly coupled with highly sensitive user behaviors and life events. 
Consider an online service whose models are periodically updated  using recent user queries. During a period of illness, a user may issue many health-related queries that are incorporated into subsequent model updates. After recovery, the user stops issuing such queries and may later request the removal of queries concerning sensitive medical conditions. 
By inferring the temporal membership, the attacker can identify the time window during which these health-related queries have been used by the model and thus reconstruct sensitive health timelines of the target user.

{\bf Adversary's Knowledge and Capabilities.} The adversary is granted black-box access to a sequence of model snapshots ${f_{1}, f_{2}, \ldots, f_{N}}$, allowing them to query each snapshot with arbitrary inputs and observe the corresponding outputs. In this work, we assume the observable output takes the format of posterior probability vectors. By default, the adversary can access all model snapshots. We also consider a more realistic setting in which the adversary has access to only a subset of the snapshots, and evaluate the robustness of our attack under this restricted access model in Section~\ref{sc:robustness}. 

Beyond black-box access, the adversary has no knowledge of the target model's internal details, including its parameters, architecture, training hyperparameters, update algorithm, or update data. In particular, the adversary does not know the contents or types of model updates (i.e., whether a sample is inserted or removed at a given update). Consequently, the temporal membership history of a target sample such as the number of membership intervals and their durations is entirely unknown to the adversary.

We further assume that the adversary is {\em passive} and cannot influence the model update process. For example, they cannot inject crafted samples into the training set and subsequently request their removal to manipulate the model's evolution. This assumption is consistent with prior work~\cite{chen2021machine,carlini2022membership} and reflects many practical deployment scenarios in which model updates are controlled solely by the service provider.

Finally, the adversary has access to one or more auxiliary shadow datasets, which need not follow the same distribution as the target model's training data. These datasets may consist of publicly available data from related domains (e.g., public image repositories or open-domain text corpora), data collected through web crawling, or auxiliary datasets released by the model provider or third parties. 

Based on the adversary's objective and capabilities, we define a {\em black-box temporal membership
inference attack} (TMIA) as learning a mapping function  $\mathcal{A}$ that takes the target sample $x$ and its outputs by a sequence of observed snapshots $\{f_{1}, f_{2}, \ldots, f_{N}\}$ and returns a set of non-overlapping time intervals during which $x$ is inferred to be a member: 
\[
\mathcal{A} :
\big(x, f_{1}(x), f_{2}(x), \ldots, f_{N}(x)\big)
\rightarrow
\{{I}_1, {I}_2, \ldots, {I}_{k}\}.
\]
By default, $x$ is considered a non-member in the model snapshots  at all timestamps not covered by the inferred intervals. 

Note that TMIA aims to reconstruct a sample's membership history over the queried snapshots only. 
Any insertion or deletion events that occur entirely between two consecutive queried snapshots are inherently unobservable to any black-box adversary, as they leave no directly observable evidence. Therefore, if a sample enters and exits the training set multiple times between two queried snapshots, these intermediate transitions cannot be recovered by TMIA without additional observations. This limitation is fundamental to the threat model rather than to our attack.

{\bf Applicability and Practical Relevance.}
TMIA can be applied to any update scenario, including streaming, online, or periodically retrained paradigms.  
The considered threat model is consistent with prior studies on privacy leakage from model updates~\cite{salem2020updates,zanella2020analyzing,jagielski2023combine,michel2026sequential}. It is also well motivated by real-world machine learning systems, where models are continuously updated to improve performance, incorporate newly collected data, or address errors and safety issues while retaining access to previous model versions.
Many machine learning platforms explicitly support versioned models that can be queried independently. For example, DataRobot~\cite{datarobot} maintains registered model versions, while H2O MLOps~\cite{h2oai} enables multiple model versions to be deployed and queried simultaneously. Likewise, commercial foundation model providers expose multiple model snapshots through their APIs. For instance, OpenAI~\cite{openai} utilizes date-based naming conventions for their API model snapshots (e.g., “gpt-4o-2024-08-06").  Therefore, an adversary can issue the same query to different model snapshots over time, making access to multiple model snapshots a realistic assumption in practice.

\subsection{Measurement of Temporal Membership} \label{sec:measurement}

Temporal membership inference seeks to determine when a particular sample has influenced the model. This setting requires privacy measurements at both timestamp and interval levels. 

{\bf Timestamp-level Measurement.} To capture the temporal membership privacy at both timestamp levels, we use a unified measurement framework that evaluates the {\em precision} and {\em recall} of the inferred membership intervals. Formally, let $\mathcal{I}^*(x)$ denote the set of ground-truth temporal membership intervals of a sample $x$, and let $\mathcal{I}(x)$ denote the set of temporal intervals inferred by the adversary.
Let $T^*(x)$ and $T(x)$ be the set of distinct membership timestamps in $\mathcal{I}^*(x)$ and $\mathcal{I}(x)$, respectively. We define {\bf timestamp-level precision (T-Pre)} as the fraction of the inferred timestamps that are correct (i.e., they belong to the ground-truth set), and \textbf{timestamp-level recall (T-Rec)} as the fraction of ground-truth timestamps that are correctly inferred:
\[\text{T-Pre(x)} = \frac{ \left| T(x) \cap T^*(x) \right| }
     { \left| T(x) \right| }, \quad \text{T-Rec}(x) =
\frac{ \left| T(x) \cap T^*(x) \right| }
     { \left| T^*(x) \right| }\]

\begin{example}
\label{example:t}
Consider the ground-truth membership interval $\mathcal{I}^*(x) =\allowbreak \{[t_1, t_4], [t_6, t_8]\}$ and the inferred interval $\mathcal{I}(x)  =\allowbreak \{[t_1, t_2], [t_5, t_8]\}$ shown in Figure \ref{figure:problem}. The corresponding timestamp sets are $T^*(x) =\allowbreak \{1, 2, 3, 4, 6, 7, 8\}$ and $T(x) =\allowbreak \{1, 2, 5, 6, 7, 8\}$. Therefore, $\text{T-Pre(x)} =\allowbreak \frac{5}{6}$, and $\text{T-Rec(x)} =\allowbreak \frac{5}{7}$.
\end{example}

We further define the \textit{timestamp-level F1 score (T-F1)} as the harmonic mean of T-Pre and T-Rec. 
A higher T-F1 score reflects a stronger membership inference capability at the timestamp level. 

{\bf Interval-level Measurement.} To assess the accuracy of inferred membership intervals, we introduce an interval-level metric that jointly evaluates interval matching and boundary precision. Since the number of predicted intervals may differ from those in the ground truth, the predicted intervals cannot be directly compared with the ground-truth ones. Therefore, we evaluate the interval-level membership inference by measuring the overlap between the interval sets. Specifically, let $\mathcal{I}^*(x)$ and $\mathcal{I}(x)$ denote the sets of ground-truth and inferred membership intervals for sample $x$, respectively. An inferred interval $I_i\in\mathcal{I}(x)$ is said to match a ground-truth interval $I^*_j\in\mathcal{I}^*(x)$ if they overlap, i.e., $I_i\cap I^*_j\neq\emptyset$. 
The \textit{interval-level precision} (I-Pre) of a predicted interval $I_i$ with respect to a matching ground-truth interval $I^*_j$ is defined as the fraction of $I_i$ that overlaps with $I^*_j$:
$\text{I-Pre}(I_i, I^*_j)
=
\frac{ \left| I_i \cap I^*_j \right| }
     { \left| I_i \right| }.$

The corresponding \textit{interval-level recall} (I-Rec) of a ground-truth interval $I^*_j$ with respect to a matching predicted  interval $I_i$ is defined as the fraction of $I^*_j$ correctly covered by $I_i$: 
$\text{I-Rec}(I^*_j, I_i)
=
\frac{ \left| I_i \cap I^*_j \right| }
     { \left| I^*_j \right| }.$

A predicted interval may match multiple ground-truth intervals, and conversely, a single ground-truth interval may be matched by multiple predicted intervals. Let $G^*(I_i)$ be the set of ground-truth intervals that match a predicted interval $I_i$, and $G(I^*_j)$ be the set of predicted intervals that match a ground-truth interval $I^*_j$.  We compute the per-interval precision and recall as follows: 
{\small
\[
\begin{aligned}
\text{I-Pre}(I_i)
&=
\frac{
\sum_{I^*_j \in G(I_i)}
\text{I-Pre}(I_i,I^*_j)
}{
|G(I_i)|
},\\
\text{I-Rec}(I^*_j)
&=
\frac{
\sum_{I_i \in G(I^*_j)}
\text{I-Rec}(I^*_j,I_i)
}{
|G(I^*_j)|
}.
\end{aligned}
\]
}
Finally, we aggregate per-interval precision and recall, and measure the \textbf{interval-level precision} and \textbf{interval-level recall} for any given sample $x$:  
{\small
\[
\begin{aligned}
\text{I-Pre}(x)
&=
\frac{
\sum_{I_i \in \mathcal{I}(x)}
\text{I-Pre}(I_i)
}{
|\mathcal{I}(x)|
},
\text{I-Rec}(x)
&=
\frac{
\sum_{I^*_j \in \mathcal{I}^*(x)}
\text{I-Rec}(I^*_j)
}{
|\mathcal{I}^*(x)|
}.
\end{aligned}
\]
}

\begin{example}
Continuing with Example \ref{example:t}, we compute $\text{I-Pre}(I_1, I^*_1) =\allowbreak \frac{2}{2} =\allowbreak 1$ and
$\text{I-Pre}(I_2, I^*_2) =\allowbreak \frac{3}{4}$, yielding $\text{I-Pre(x)} =\allowbreak \frac{7}{8}$. Similarly, $\text{I-Rec}(I_1, I^*_1) =\allowbreak \frac{1}{2}$ and
$\text{I-Rec}(I_2, I^*_2) =\allowbreak \frac{3}{3} =\allowbreak 1$, resulting in $\text{I-Rec}(x) =\allowbreak \frac{3}{4}$.
\end{example}

Finally, we compute the {\bf interval-level F1 score (I-F1)} as the harmonic mean of I-Pre and I-Rec values. 
A larger I-F1 score indicates higher interval-level membership privacy risk. 

Together, timestamp-level and interval-level metrics provide complementary views of the inferred temporal intervals. Timestamp-level metrics measure fine-grained, point-wise detection performance, whereas interval-level metrics evaluate whether the detected timestamps are properly grouped into coherent temporal segments.  

Note that a prediction can achieve high timestamp-level accuracy yet low interval-level accuracy if correct timestamps are fragmented into many small intervals, because I-Pre and I-Rec are averaged over intervals and therefore penalize fragmented predictions. Conversely, high interval-level accuracy with low timestamp-level accuracy is unlikely, since significant timestamp-level errors typically reduce interval overlap and negatively affect interval-level performance.

\section{Motivation of Attack Design} 
\label{sec-motivation}

\subsection{Characterization of Temporal Membership Signals}
\label{sc:characterization}

Continual model updates fundamentally alter model behavior over time. Understanding these behavioral changes is critical for revealing when and how membership signals emerge, and for informing the design of effective temporal membership inference attacks. To this end, we first conduct an empirical analysis to characterize how model updates influence black-box model behaviors, particularly confidence trajectories, and to identify consistent signatures associated with true data membership changes.

In our empirical study, we train two models: Pythia-70M model \cite{biderman2023pythia} on the SST-5 dataset and SimpleCNN~\cite{chen2021machine} on the CIFAR-10 dataset\footnote{Details of the dataset and model are provided in Appendix~\ref{appendix:setup}.}.
At each update step, we randomly sample 1\% the original training data as the removal set, along with a set of in-distribution samples not included in the training data of the same amount as the insertion set. We adopt fine-tuning as the update algorithm. %
Finally, we randomly select a target sample from the training data, which is removed and subsequently re-inserted at a later model update, and track its predicted confidence across successive model snapshots.

\begin{figure}[!t]
	\begin{center}
    \includegraphics[width=0.43\textwidth]{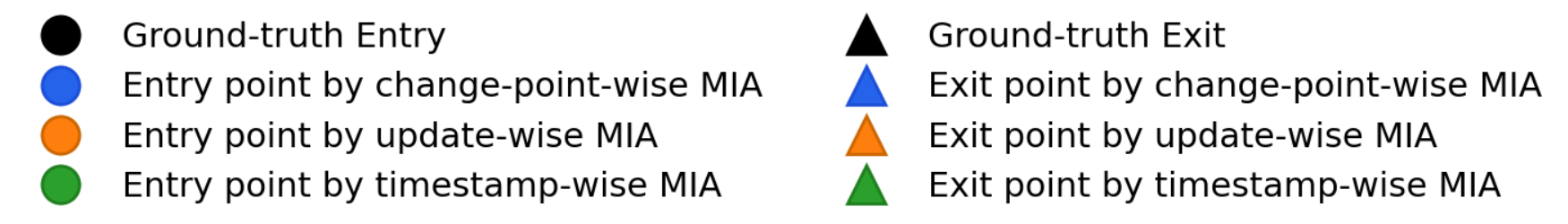}
    
        \includegraphics[width=0.35\textwidth]{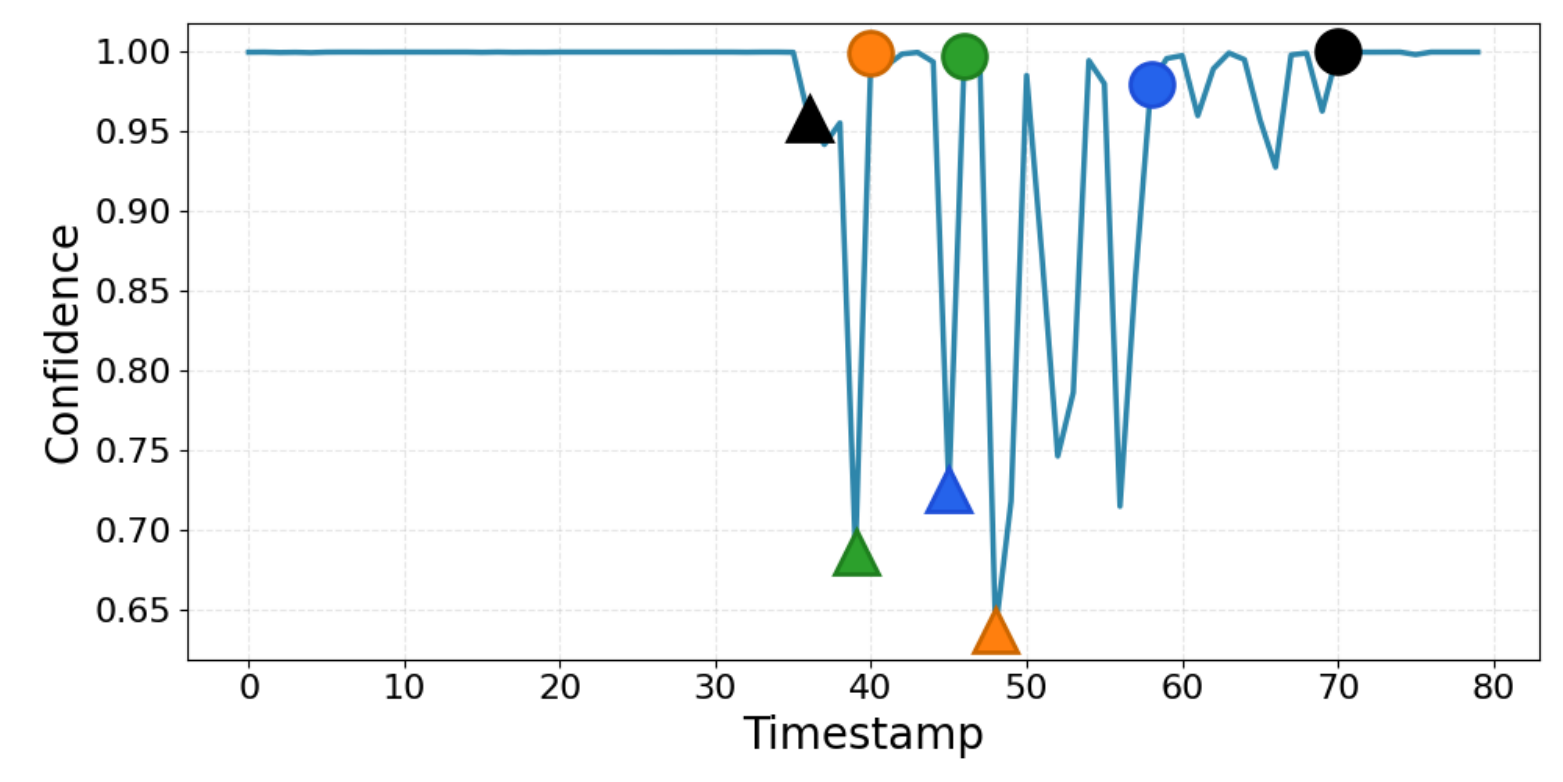}
        \label{fig:SST5-two-insert-motivation}
		\caption{Entry and exit points predicted by three straw-man approaches (Pythia-70M model, SST-5 datasets). The target sample was initially present in the training set. 
        It was removed at the 36th timestamp, and re-inserted at the 70th timestamp. }
        \vspace{-0.1in}
        \label{figure:conf_timeseries}
	\end{center}
\end{figure}

Figure~\ref{figure:conf_timeseries} illustrates the confidence score of the target sample across 80 snapshots of the Pythia-70M model. The results of the SimpleCNN model are similar and can be found in Appendix \ref{appendix:emp-charactirization}. 
We observe a pronounced disparity between insertion and deletion in their effects on a sample’s confidence score. While both operations alter the model’s outputs, insertion results in rapid stabilization: once the sample is incorporated into the training set, its confidence quickly converges and remains relatively stable under subsequent updates. In contrast, deletion exhibits a more persistent influence, with the sample’s confidence continuing to fluctuate as the model undergoes further updates.
This disparity indicates that the temporal dynamics of confidence scores can be leveraged not only to identify the timing of model updates but also to distinguish between sample insertion and removal events. 

\subsection{Straw-man Approaches and Their Insufficiency}
\label{sc:strawman}
Can existing approaches be readily adapted to our setting to identify the entry and exit points of a target sample? We investigate this question by considering three straw-man baselines: \textit{timestamp-wise MIA (TS-MIA)}, \textit{update-wise MIA (UP-MIA)}, and \textit{change-point–wise MIAs (CP-MIA)}.
TS-MIA performs membership inference independently at each timestamp, UP-MIA relies on pairwise comparisons between consecutive model snapshots, and CP-MIA applies statistical analysis over the entire sequence of model snapshots. 
We describe the details of each approach below. 
To highlight their limitations, we evaluate these attacks under the same experimental setting as described in Section~\ref{sc:characterization} and visualize the entry and exit points detected by each method in Figure~\ref{figure:conf_timeseries}. To rule out inaccuracies arising from uncertainty in the number of update events and to provide a strong baseline for comparison, we further assume that the adversary underlying all three straw-man approaches knows the total number of update events. 

\smallskip
{\bf Timestamp-wise MIA (TS-MIA).}
TS-MIA applies an existing MIA developed for static models~\cite{shokri2017membership,song2021systematic} independently to each model snapshot, aiming to determine whether a given sample is present in that snapshot. As shown in Figure~\ref{figure:conf_timeseries}, by operating on each snapshot in isolation, TS-MIA discards valuable temporal dependencies, resulting in poor accuracy when identifying entry and exit points. For instance, it infers an entry point at the 46th timestamp, which is substantially off from the ground truth at the 70th timestamp. 

\smallskip
{\bf Update-wise MIA (UP-MIA).}
We consider the \textit{delta attack}~\cite{jagielski2023combine}, a state-of-the-art MIA against model updates. It classifies a sample as a member of the update set at time point $t_i$ if it exhibits a large loss difference between two consecutive model snapshots,$f_{i-1}$ and $f_i$. 

One major weakness of the delta attack is that relying solely on adjacent snapshots provides insufficient temporal context for accurate temporal membership inference. As shown in Figure~\ref{figure:conf_timeseries}, even when the target sample is not included in any update set between the 36th and 70th timestamps, its confidence score still fluctuates substantially across consecutive model snapshots. Importantly, the magnitude of these fluctuations can be comparable to---or even exceed---those induced by true update events. Because UP-MIA  infers membership solely from the maximum output change between adjacent model snapshots, it can be severely misaligned with the true insertion and deletion timestamps, resulting in large deviations from the ground truth (e.g., incorrectly identifying the 40th timestamp as an entry point in Figure~\ref{figure:conf_timeseries}). Even worse, UD-MIA can infer an incorrect temporal ordering of update events by predicting the exit point after the entry point, indicating a wrong insertion-then-removal update event. 

\smallskip
{\bf Change-point-wise MIA (CP-MIA).} 
Intuitively, the model outputs across all snapshots can be treated as a time-series sequence, and the inference attack can be framed as a change-point detection problem whose goal is to identify the precise times at which these statistical shifts occur~\cite{aminikhanghahi2017survey}. Following this idea, we apply \textit{Cumulative Sum (CUSUM)}~\cite{kleiber2017structural}, a state-of-the-art change-point detection method that detects abrupt changes in the mean or variance of a time series.
Since the original CUSUM method does not distinguish between entry and exit points, we introduce an additional constraint to make CP-MIA a strong baseline: the adversary is assumed to know the temporal order of update events. 

Figure~\ref{figure:conf_timeseries} shows that the change points identified by CP-MIA do not align with the true entry and exit points of the target sample. 
For instance, it predicts an exit point at the 45th timestamp, which is considerably distant from the ground truth at the 36th timestamp. 
This highlights a fundamental limitation of CP-MIA: statistical shifts in the time series alone are insufficient to accurately infer temporal membership, even under strong adversarial knowledge of the update sequence.

To summarize, our empirical results demonstrate that the dynamics of model outputs cannot be adequately captured by timestamp-wise membership reasoning (TS-MIA), pairwise snapshot comparisons (UP-MIA), or sequence-wise statistical analysis (CP-MIA). This highlights the need for designing effective attacks that explicitly update-induced model dynamics.

\section{Attack Methodology}
\label{sec-Method}

In this section, we present the design of \system, our proposed black-box temporal membership inference attack. \system\ infers temporal membership intervals by analyzing model updates across multiple timestamps. The framework consists of three main components: \emph{training update-aligned shadow models}, \emph{labeling timestamps}, and \emph{constructing membership intervals}. Below, we describe each phase in detail.

\subsection{Training Update-Aligned Shadow Models}
\label{sc:shadow}
Recall that the adversary has only black-box access to the target model. To approximate the behavior of the target model under continuous updates, the adversary constructs a sequence of shadow models that emulate the target model’s evolution over time. 
Formally, let $\{f_i\}_{i=1}^N$ denote the sequence of target model snapshots. 
The adversary’s goal is to construct a corresponding sequence of shadow models $\{f_i^S\}_{i=1}^N$ such that these shadow models approximate the update behavior of $\{f_i\}_{i=1}^N$. 
This alignment enables the adversary to capture temporal changes in the model’s outputs and leverage them for temporal membership inference in the next steps.

Unlike conventional MIAs \cite{shokri2017membership,salem2018ml,yeom2018privacy} that train one or more static shadow models to approximate a single fixed target model, training update-aligned shadow models introduces several difficulties. 
First, the shadow model must mimic the target model’s behavior across successive updates, despite that the adversary has no knowledge of the update algorithm used by the target model. 
Second, continual updates introduce distribution shift in the model output, making it difficult for shadow models to emulate the target model’s update behavior across all time steps.
These challenges make update-aligned shadow model training  more complex than static shadow model training.

To train the update-aligned shadow models, the adversary first trains an initial shadow model
$f^S_0$, using a shadow dataset $D_S$ sampled from the adversary’s
auxiliary knowledge. We assume that the architecture, dataset distribution and hyperparameters of the shadow
model may differ from those of the target model. 
The adversary then performs a sequence of model updates, yielding  a series of shadow models
$\{f^S_1, f^S_2, \ldots, f^S_N\}$. 
At each timestamp $t_i$, the adversary simulates the target model’s update by updating $f^S_{i-1}$ with a \textit{shadow update set} $\Delta^S_i$. 
The shadow update sets need not follow the same distribution as the target model's update sets. 
Since the adversary 
does not have the knowledge of the update algorithm  used by the target model either, there can be mismatch between the updates by the target model and the shadow model. We will empirically evaluate the
effectiveness of \system\ under such discrepancies in Section \ref{sec-Evaluation}.

\subsection{Labeling Timestamps}

We define the task of identifying the entry and exit points as a sequence labeling problem. Formally, given a sample $x$ and its corresponding time-series sequence $\s=(s_1, s_2, \ldots, s_N)$, where each $s_i$ denotes the output of model snapshot $f_i$ on $x$, i.e., $s_i = f_i(x)$, the goal is to learn a mapping function $M$ that maps each timestamp in $s$ with a specific label:
\[
M: s_i \rightarrow \{\mathsf{N},\mathsf{E},\mathsf{X}\}, \quad \forall s_i\in \s,
\]
where label~$\mathsf{N}$ indicates that the timestamp $i$ is a non-change point of $x$; label~$\mathsf{E}$ indicates that timestamp~$i$ is an entry point of $x$ (i.e., $x$ is inserted into $f_i$); and label~$\mathsf{X}$ indicates that timestamp~$i$ is an exit point of $x$ (i.e., $x$ is removed from $f_i$).

We train the sequence labeling model by two steps: constructing a training dataset, and training an LSTM-based model. Next, we explain the details of each step. 

\subsubsection{Constructing Sequence Labeling Training Dataset}
After the adversary obtained a sequence of update-aligned shadow models $\{f^S_0, f^S_1, f^S_2, \ldots, f^S_N\}$ (Section \ref{sc:shadow}), the adversary randomly selects a set of samples $X_S$, each sampled from the sets $\{D_S, \Delta^S_1, \dots, \Delta^S_N\}$ according to a uniform distribution. Here, $D_S$ denotes the training set used to train $f^S_0$, and $\Delta_i^S$ is the update set used to update the model snapshot $f^S_{i}$. Next, for each sample $x\in X_S$, the adversary queries each shadow model with $x$ 
and collects the output posterior probability vectors, forming a time series sequence $\big(
f^S_0(x), f^S_1(x), \ldots, f^S_N(x)
\big)$. 
We obtain a univariate time series $\s = (s_1, s_2, \ldots, s_N)$ accordingly, where $s_i$ denotes the confidence score extracted from $f^S_i(x)$. 
This time series captures the temporal evolution of the
model’s response to the target sample across updates.
After obtaining the time series $\s$, the adversary assigns each value $s_i\in \s$ a label from $\{\mathsf{N},\mathsf{E},\mathsf{X}\}$, depending on whether $x$ is a non-change point, an entry point, or an exit point in the model snapshot $f^S_i$, respectively. 

\subsubsection{Training Sequence Labeling Model}
\label{sc:slm}

\begin{figure}[t!]
	\begin{center}
  \includegraphics[width=0.85\linewidth]{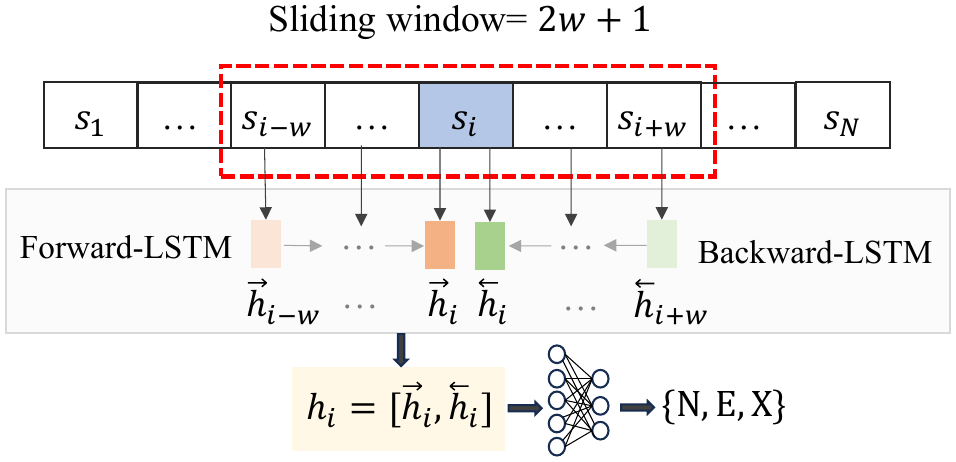}
		\caption{Illustration of the sequence labeling model. }
        \vspace{-0.2in}
        \label{figure:attack_training}
	\end{center}
\end{figure}

We train a LSTM model for sequence labeling, since they capture temporal dependencies across time steps. 
However, as traditional LSTM models are primarily designed to capture global long-term dependencies, they may fail to detect membership transition events as the incorporated distant temporal context may become noise to the inference. 
To address this issue, we adopt the sliding-window based LSTM approach~\cite{atashgahi2022memory,peng2023mad,shah2021sequence} to our setting, and design a \textit{a bidirectional LSTM} (BiLSTM) \cite{graves2005framewise} incorporated with  a sliding window, aiming to emphasize local temporal patterns around entry and exit points. Figure~\ref{figure:attack_training} illustrates our sequence labeling  model. 

Specifically, given a time series $\mathbf{s}$, for each timestamp $t_i$ in $\mathbf{s}$, we construct a fixed-length \emph{sliding window} $\mathbf{c}_i$ of size $2w+1$: $\mathbf{c}_i = (t_{i-w}, \ldots, t_i, \ldots, t_{i+w})$. Near the beginning or end of the sequence, where the sliding window extends beyond the time series, we pad all out-of-range window positions by replicating the nearest boundary value ($s_1$ or $s_N$).

Next, we incorporate both forward and backward LSTM in the BiLSTM model with the sliding window, ensuring the hidden state at time $t_i$ only depends on inputs from $t_{i-w}$ to $t_i$, not from $t_1$ to $t_i$. Specifically, let the effective start of the window at time $t_i$ be 
\begin{equation}
\label{eqn:window-start}
s(i) = \max(1,\, i - w).
\end{equation}

The forward LSTM update is then defined as
\begin{equation}
\label{eqn:lstm-forward-window}
(\overrightarrow{\mathbf{h}}_{i}, \overrightarrow{\mathbf{c}}_{i})
=
\begin{cases}
\mathrm{LSTM}_{\mathrm{fwd}}
\bigl(\mathbf{x}_{i},\; \mathbf{h}_0,\; \mathbf{c}_0 \bigr),
& \text{if } i = s({i}), \\[6pt]
\mathrm{LSTM}_{\mathrm{fwd}}
\bigl(\mathbf{x}_{i},\; \overrightarrow{\mathbf{h}}_{i-1},\; \overrightarrow{\mathbf{c}}_{i-1} \bigr),
& \text{if } i > s(i),
\end{cases}
\end{equation}
where $\mathbf{h}_0,\; \mathbf{c}_0$ are reset (e.g., zero-initialized or learned) at the beginning of each sliding window, and $\mathrm{LSTM}_{\mathrm{fwd}}(\cdot)$ is defined in Eqn. \eqref{eqn:lstm}.

Similarly, let the effective end of the window at time $t_i$ be 
\begin{equation}
\label{eqn:window-end}
e(i) = \min(N,\, i + w).
\end{equation}
Here, $N$ is the number of timestamps in the whole sequence. 

The backward LSTM update is then defined as the follows:

\begin{equation}
\label{eqn:lstm-backward-window}
(\overleftarrow{\mathbf{h}}_{i}, \overleftarrow{\mathbf{c}}_{i})
=
\begin{cases}
\mathrm{LSTM}_{\mathrm{bwd}}
\bigl(\mathbf{x}_{i},\; \mathbf{h}_0,\; \mathbf{c}_0 \bigr),
& \text{if } {i} = e({i}), \\[6pt]
\mathrm{LSTM}_{\mathrm{bwd}}
\bigl(\mathbf{x}_{i},\; \overleftarrow{\mathbf{h}}_{i+1},\; \overleftarrow{\mathbf{c}}_{i+1} \bigr),
& \text{if } i < e(i),
\end{cases}
\end{equation}
Here, $\mathrm{LSTM}_{\mathrm{bwd}}(\cdot)$ is defined in Eqn. \eqref{eqn:lstm}. The hidden and cell states are reset at the end of each window, so that the hidden state at time $t_i$ depends only on the inputs from $t_i$ to $t_{i+w}$ within the sliding window, and is independent of inputs beyond the window.

Finally, the hidden states at timestamp $t_i$ is obtained by concatenating  the forward and backward hidden states:
\begin{equation}
\label{eqn:hk}
\mathbf{h}_i
=
\big[
\overrightarrow{\mathbf{h}}_{i} \,;\,
\overleftarrow{\mathbf{h}}_{i}
\big],
\end{equation}

After obtaining $\mathbf{h}_i$, it is passed through a linear layer followed by a softmax activation to produce class probabilities. We use cross-entropy as the loss function. In this way, the model maps the learned representations to the target labels.

\subsection{Constructing Membership Intervals}
\label{subsec:refinement}

A major limitation of the sequence labeling model is that assigning labels to individual timestamps independently fails
to capture the temporal dependencies between entry and exit points. The model may predict an exit point before the corresponding entry point, or generate isolated entry or exit labels in the sequence. Such inconsistencies can prevent the construction of valid membership intervals. 

First, we formally define {\em valid membership intervals}.

\begin{definition}[Valid Membership Intervals]
\label{def:valid-intervals}
Let $T=\allowbreak ((t_1,y_1),\allowbreak (t_2,y_2),\allowbreak\ldots,\allowbreak (t_N,y_N))$ be a sequence of timestamps, where each timestamp $t_i$ is assigned a label
$y_i\in\{\mathsf{N},\mathsf{E},\mathsf{X}\}$ by our BiLSTM model, 
representing a \emph{non-change}, \emph{entry}, and \emph{exit} point, respectively. Let $E(T)=\{i\mid y_i=\mathsf{E}\}$, and
$X(T)=\{i\mid y_i=\mathsf{X}\}$ denote the sets of entry and exit timestamps, respectively. 
A set of membership intervals is \emph{valid} if there exists an injective matching function: $\phi:X(T)\rightarrow E(T)$ that satisfies the following three requirements:

\begin{enumerate}
    \item \textbf{Correct ordering.}
    Every exit point is matched to a preceding entry point:
    $\forall\,j\in X(T)$, $\phi(j)<j.$

    \item \textbf{Entry-exit matching.}
    Every exit point in $X(T)$ is matched to exactly one entry point in $E(T)$, while each entry point $E(T)$ is matched to at most one exit point in $X(T)$. Consequently, an entry point may remain unmatched, corresponding to a membership interval that extends to the last observed timestamp $t_N$.

    \item \textbf{Non-overlapping intervals.}
    For any two intervals $I_i$, $I_j$ induced by $\phi$, $I_i\cap I_j=\emptyset$. 
\end{enumerate}
\end{definition}

To construct valid temporal
membership intervals, we design a three-step procedure:
Step~1 extracts the entry and exit points and partitions them
into groups, Step~2 selects a representative timestamp for
each group, and Step~3 pairs the representatives to form
membership intervals. Step~1 ensures Requirement 1, while Steps 2 \& 3 ensures Requirements 2 \& 3. of valid membership intervals to be satisfied. Next, we discuss the details of each step. 

{\bf Step 1:} We discard all non-change points and keep  all entry and exit timestamps in $\mathcal{E}$, sorted in increasing order. 
We partition $\mathcal{E}$ into maximal groups of consecutive same-label
timestamps $\mathcal{G}=\{\mathcal{G}_1,\ldots,\mathcal{G}_K\}$. Each
group $\mathcal{G}_q = \{e_a, e_{a+1}, \ldots, e_b\}$ is a maximal run  with
$\ell_{e_a} = \ell_{e_{a+1}} = \cdots = \ell_{e_b}$, and
$\ell_{e_{a-1}} \neq \ell_{e_a}$, $\ell_{e_{b+1}} \neq \ell_{e_b}$.

{\bf Step 2:} For each group $\mathcal{G}_q$, we select the timestamp with the highest confidence score as the \textit{representative timestamp} $\tau_q$.  
Let
$\boldsymbol{\tau}=(\tau_1,\ldots,\tau_K)$
denote the chronologically ordered sequence of representative
timestamps.

{\bf Step 3:} We process $\boldsymbol{\tau}$ in chronological order. For each
$\tau_k$ with $\ell_{\tau_k}=\mathsf{X}$, we match it to the immediately preceding
unmatched entry $\tau_j<\tau_k$ ($\ell_{\tau_j}=\mathsf{E}$); by alternation this is the
entry right before it. Each matched entry--exit pair $(\tau_j,\tau_k)$ defines the
closed membership interval
$
[t_{\tau_j},t_{\tau_k-1}],
$
There are two boundary cases:
(i) if the sequence ends with an unmatched entry $\tau_K$,
the sample is inferred to remain a member through the final
snapshot, giving $[t_{\tau_K},t_N]$; and
(ii) if the sequence begins with an exit $\tau_1$, 
$x$ is inferred to be included in the initial training set and use the implicit
entry point $t_0$, giving $[t_0,t_{\tau_1-1}]$. 

\begin{figure}[t!]
	\begin{center}
  \includegraphics[width=0.8\linewidth]{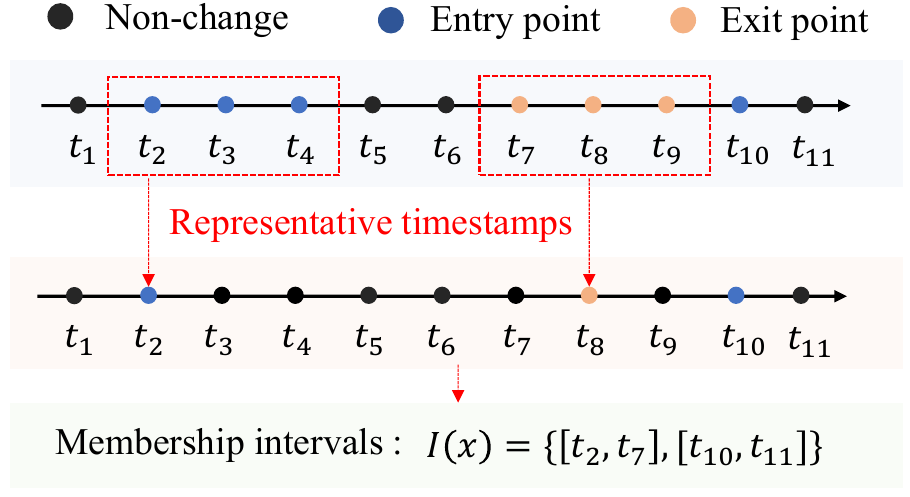}
		\caption{An example of membership interval construction. }
        \label{figure:inference}
        \vspace{-0.2in}
	\end{center}
\end{figure}

Figure~\ref{figure:inference} illustrates an example of our membership interval construction process. Building on this construction process, we formally state the following theorem, which establishes the correctness of our method. 

\begin{theorem}
    \label{theorm:validinterval}
    The three-step process guarantees to return valid membership intervals (Definition \ref{def:valid-intervals}). 
\end{theorem}

The proof of Theorem \ref{theorm:validinterval} can be found in Appendix~\ref{appendix:proof-theorem}.  By guaranteeing valid membership intervals, Theorem \ref{theorm:validinterval} formalizes the correctness of our approach and provides a foundation for interval-level measurement. %

\section{Evaluation}
\label{sec-Evaluation}

This section presents the results of our empirical evaluation. 
All the experiments are executed on a server with 4 NVIDIA A100
GPUs, each with 40GB of memory. 
All the experimental results are obtained as the average of three trials.

\subsection{Setup}

\textbf{Models and Datasets}.
To assess the generality of our approach across different modalities and architectures, we consider both language and vision models in our evaluation: (1) {\em Language models} (LM): we consider three widely-used transformer-based architectures: \textit{Pythia-70M}~\cite{biderman2023pythia}, \textit{T5}~\cite{raffel2020exploring}, and \textit{GPT-2 Small}~\cite{radford2019language}; 
(2) Image classification DNNs: \textit{ResNet-18}~\cite{he2016deep}, \textit{SimpleCNN}~\cite{chen2021machine}, and \textit{MobileNet}~\cite{howard2017mobilenets}. 
We use three widely-adopted text datasets:  
\textit{SST-5}~\cite{socher2013recursive}, \textit{News20}~\cite{albishre2015effective}, \textit{MNLI}~\cite{wang2018glue}, and three image benchmark datasets: \textit{CIFAR-10}~\cite{krizhevsky2009learning}, \textit{SVHN}~\cite{netzer2011reading}, and \textit{CINIC-10}~\cite{darlow2018cinic}. 

\textbf{Updates.} 
We simulate 80 and 60 update steps for the examined LMs and DNNs, respectively. Each update includes both insertions and removals.
The update sizes are dataset-dependent: 1\% for SST-5 and News20, and 0.5\% for MNLI.
All inserted and removed samples are randomly drawn from the dataset. Under this update process, a sample may be inserted into and removed from the training set multiple times throughout the update sequence. Across datasets, 31\%--55\% of updated samples have at least two membership intervals, with each sample undergoing up to three or four insertion/removal events depending on the dataset. Additional details of update sampling, ratio calculation, and evaluation-sample composition are provided in Appendix~\ref{appendix:data-updates}.  
Beyond this default setting, we evaluate \system\ under correlated updates, model quantization, and partial observation of model snapshots (Section~\ref{sc:robustness}).

Regarding model update algorithms, we consider three algorithms for LMs: \textit{fine-tuning}~\cite{maini2024tofu} for both data insertion and removal, as well as \textit{Gradient Ascent (GA)}~\cite{jang2023knowledge}  and \textit{Negative Preference Optimization (NPO)}~\cite{zhang2024negative} for data removal (forgetting). For DNNs, besides fine-tuning and GA, 
we consider \textit{SCRUB}~\cite{kurmanji2023towards} for data forgetting. The implementation details of these algorithms are described in Appendix~\ref{appendix:unlearning-methods}. 

\textbf{Evaluation Metrics}.  
We evaluate timestamp-level inference accuracy using T-Pre, T-Rec, and T-F1, and interval-level inference accuracy using I-Pre, I-Rec, and I-F1 (Section \ref{sec:measurement}). Higher T-F1 and I-F1 scores indicate stronger attack effectiveness. Since our attack directly predicts temporal membership intervals rather than producing probabilistic or threshold-based decisions, AUC is not an applicable evaluation metric.

\textbf{Compared Methods\footnote{The CP-MIA method assumes a known number of membership transitions and is therefore not directly included in the comparison.}}.  
We consider two approaches: \ding{172}  \textbf{Timestamp-wise MIA (TS-MIA)}: We apply the \textit{entropy-based MIA}\footnote{https://github.com/inspire-group/membership-inference-evaluation}~\cite{song2021systematic}, one of the recent benchmark MIAs, independently to each model snapshot; and 
\ding{173} 
 {\bf Update-wise MIA (UP-MIA)}: We adapt the \textit{delta attack}~\cite{jagielski2023combine}, the only prior update-wise MIA, to our setting. The attack compares outputs of consecutive model snapshots to infer entry and exit points: round~$i$ is an entry if the loss difference exceeds a threshold, and an exit if it falls below. Thresholds are set using the shadow-model calibration strategy from \cite{jagielski2023combine}. 

More details of the experimental setup, including the datasets, target models, data updates, attack model and its training process, can be found in Appendix \ref{appendix:setup}. 

\subsection{Performance of \system\ }
\definecolor{light-gray}{gray}{0.8}

\begin{table*}[!t]
\centering
\caption{Timestamp-level accuracy of \system\ and baselines.  The best T-F1 results for each setting are marked in \gray{gray}. }
\label{tab:TL-performance}
\begin{tabular}{c|c|c|c|c|c|c|c|c|c|c}
\hline
\multirow{2}{*}{\textbf{Model}} & \multirow{2}{*}{\textbf{Dataset}} & \multicolumn{3}{c|}{\textbf{TMIA (Ours)}} & \multicolumn{3}{c|}{\textbf{Timestamp-wise MIA}} & \multicolumn{3}{c}{\textbf{Update-wise MIA}} \\ \cline{3-11} 
 &  & \textbf{T-Pre} & \textbf{T-Rec} & \textbf{T-F1} & \textbf{T-Pre} & \textbf{T-Rec} & \textbf{T-F1} & \textbf{T-Pre} & \textbf{T-Rec} & \textbf{T-F1} \\ \hline
\multirow{3}{*}{Pythia-70M}     
& SST-5   & 88.78 & 94.06 & \gray{91.34} & 69.26 & 88.21 & 77.59 & 52.01 & 65.25 & 57.88 \\
& News20  & 82.55 & 84.58 & \gray{83.55} & 57.33 & 79.44 & 66.60 & 29.01 & 59.81 & 39.07 \\
& MNLI    & 83.67 & 91.19 & \gray{87.27} & 72.91 & 94.16 & 82.18 & 45.35 & 75.41 & 56.64 \\ \hline
\multirow{3}{*}{T5}             
& SST-5   & 84.73 & 92.19 & \gray{88.30} & 61.53 & 90.94 & 73.40 & 46.35 & 89.87 & 61.16 \\
& News20  & 80.51 & 88.10 & \gray{84.13} & 57.25 & 81.31 & 67.19 & 31.78 & 63.26 & 42.30 \\
& MNLI    & 90.32 & 91.92 & \gray{91.10} & 64.79 & 91.77 & 75.95 & 42.88 & 75.26 & 54.64 \\ \hline
\multirow{3}{*}{GPT-2}          
& SST-5   & 83.54 & 90.52 & \gray{86.89} & 60.96 & 91.01 & 73.01 & 38.02 & 78.30 & 51.18 \\
& News20  & 82.64 & 84.88 & \gray{83.75} & 52.65 & 85.19 & 65.08 & 30.71 & 61.80 & 41.03 \\
& MNLI    & 81.28 & 86.79 & \gray{83.94} & 68.25 & 86.33 & 76.23 & 41.63 & 71.01 & 52.49 \\ \hline
\multirow{3}{*}{ResNet-18}      
& CIFAR10 & 77.33 & 81.48 & \gray{79.35} & 59.51 & 84.73 & 69.91 & 39.53 & 73.28 & 51.36 \\
& SVHN    & 71.30 & 74.24 & \gray{72.74} & 43.23 & 81.76 & 56.55 & 33.77 & 59.94 & 43.20 \\
& CINIC10 & 77.06 & 83.75 & \gray{80.27} & 60.54 & 88.36 & 71.85 & 44.83 & 68.11 & 54.07 \\ \hline
\multirow{3}{*}{SimpleCNN}     
& CIFAR10 & 80.22 & 76.39 & \gray{78.26} & 66.40 & 85.78 & 74.85 & 35.00 & 75.07 & 47.74 \\
& SVHN    & 70.43 & 72.69 & \gray{71.02} & 47.18 & 86.68 & 61.10 & 34.56 & 70.48 & 46.38 \\
& CINIC10 & 78.28 & 76.87 & \gray{77.57} & 66.74 & 85.49 & 74.96 & 51.13 & 79.35 & 62.19 \\ \hline
\multirow{3}{*}{MobileNet}      
& CIFAR10 & 80.36 & 76.86 & \gray{78.57} & 66.17 & 87.81 & 75.47 & 46.26 & 79.19 & 58.40 \\
& SVHN    & 72.15 & 77.38 & \gray{74.23} & 41.90 & 96.04 & 58.35 & 34.07 & 65.64 & 44.85 \\
& CINIC10 & 81.34 & 73.65 & \gray{77.30} & 64.77 & 87.68 & 74.50 & 57.54 & 79.61 & 66.80 \\ \hline
\end{tabular}
\end{table*}

\begin{table*}[!t]
\centering
\caption{Interval-level accuracy of \system\ and baselines.  The best I-F1 results for each setting are marked in \gray{gray}. }
\label{tab:IL-performance}
\begin{tabular}{c|c|c|c|c|c|c|c|c|c|c}
\hline
\multirow{2}{*}{\textbf{Model}} & \multirow{2}{*}{\textbf{Dataset}} & \multicolumn{3}{c|}{\textbf{TMIA (Ours)}} & \multicolumn{3}{c|}{\textbf{Timestamp-wise MIA}} & \multicolumn{3}{c}{\textbf{Update-wise MIA}} \\ \cline{3-11} 
 &  & \textbf{I-Pre} & \textbf{I-Rec} & \textbf{I-F1} & \textbf{I-Pre} & \textbf{I-Rec} & \textbf{I-F1} & \textbf{I-Pre} & \textbf{I-Rec} & \textbf{I-F1} \\ \hline
\multirow{3}{*}{Pythia-70M}     
& SST-5   & 80.89 & 87.70 & \gray{84.16} & 56.47 & 57.57 & 57.01 & 46.93 & 68.35 & 55.65 \\
& News20  & 75.04 & 79.58 & \gray{77.24} & 47.52 & 50.85 & 49.13 & 30.66 & 64.35 & 41.53 \\
& MNLI    & 78.84 & 76.63 & \gray{77.72} & 46.50 & 59.73 & 52.29 & 39.16 & 74.89 & 51.43 \\ \hline
\multirow{3}{*}{T5}             
& SST-5   & 68.70 & 77.97 & \gray{73.04} & 53.23 & 61.25 & 56.96 & 29.85 & 87.63 & 44.53 \\
& News20  & 74.15 & 71.69 & \gray{72.90} & 50.31 & 51.92 & 51.10 & 35.70 & 68.67 & 46.98 \\
& MNLI    & 72.81 & 69.61 & \gray{71.17} & 43.59 & 57.04 & 49.42 & 39.62 & 74.64 & 51.76 \\ \hline
\multirow{3}{*}{GPT-2}          
& SST-5   & 69.20 & 82.41 & \gray{75.23} & 47.00 & 64.03 & 54.21 & 31.17 & 80.45 & 44.93 \\
& News20  & 75.13 & 77.66 & \gray{76.37} & 39.41 & 58.81 & 47.19 & 34.35 & 68.29 & 45.71 \\
& MNLI    & 76.38 & 63.50 & \gray{69.35} & 49.18 & 38.68 & 43.30 & 34.89 & 70.66 & 46.71 \\ \hline
\multirow{3}{*}{ResNet-18}      
& CIFAR10 & 74.75 & 68.26 & \gray{71.36} & 40.74 & 43.02 & 41.85 & 23.93 & 67.20 & 35.29 \\
& SVHN    & 63.47 & 67.94 & \gray{65.63} & 26.23 & 36.20 & 30.42 & 34.59 & 36.21 & 35.38 \\
& CINIC10 & 75.01 & 59.07 & \gray{66.09} & 41.55 & 52.09 & 46.23 & 35.35 & 72.55 & 47.54 \\ \hline
\multirow{3}{*}{SimpleCNN}     
& CIFAR10 & 79.05 & 53.26 & \gray{63.64} & 47.56 & 40.65 & 43.83 & 23.26 & 50.30 & 31.81 \\
& SVHN    & 73.76 & 56.28 & \gray{63.85} & 29.95 & 36.67 & 32.97 & 33.40 & 27.89 & 30.40 \\
& CINIC10 & 70.10 & 62.63 & \gray{66.15} & 50.29 & 44.21 & 47.05 & 29.89 & 65.15 & 40.98 \\ \hline
\multirow{3}{*}{MobileNet}      
& CIFAR10 & 80.16 & 62.36 & \gray{70.15} & 46.49 & 43.61 & 45.00 & 26.34 & 61.00 & 36.79 \\
& SVHN    & 71.36 & 63.04 & \gray{66.94} & 20.14 & 42.79 & 27.39 & 33.56 & 34.44 & 33.99 \\
& CINIC10 & 75.62 & 60.35 & \gray{67.13} & 45.45 & 44.51 & 44.98 & 36.66 & 72.19 & 48.63 \\ \hline
\end{tabular}
\end{table*}

In this section, we report the timestamp-level and interval-level inference performance of \system\ and the baseline methods. In addition, we measure the accuracy of inferred entry/exit points, and report the results in Appendix~\ref{appendix:inference-entry-exit}.  

{\bf Timestamp-level Inference.} 
Table~\ref{tab:TL-performance} reports the T-Pre, T-Rec, and T-F1 of \system\ and the two baselines. \system\ achieves consistently strong T-Pre, T-Rec, and T-F1 results across all the settings. 
Specifically, across the three language models, \system\ achieves both high T-Pre and T-Rec results ($\geq 80.51\%$), with T-F1 score higher than 83\% in all settings. 

Compared with the baselines, \system\ significantly outperforms both in terms of T-F1 score across all the settings. Specifically, while TS-MIA achieves high T-Rec (often exceeding 85\%), it does so by predicting entry points at nearly all timestamps for all test samples. This behavior leads to low precision, with T-Pre typically remaining below 70\%, which significantly limits its overall T-F1 performance.
In contrast, UP-MIA consistently yields low T-Pre and T-Rec values across different models and datasets, resulting in substantially lower T-F1 scores. These results demonstrate that \system\ achieves a significantly better balance between precision and recall, enabling more accurate timestamp-level membership inference intervals in all setting. 

{\bf Interval-level Inference.}
Table~\ref{tab:IL-performance} reports the I-Pre, I-Rec, and I-F1 results of \system\ and the two baselines. \system\ also achieves consistently strong interval-level attack performance on all models and datasets, yielding favorable I-Pre, I-Rec, and I-F1 results. For example, the I-F1 score of \system\ exceeds 72\% when attacking the Pythia-70M model across the three text datasets.
Furthermore, \system\ consistently attains higher I-Pre and I-Rec than the baselines in most settings, and achieves substantially higher I-F1 across all cases, with observed relative improvements of at least 21.82\%.
These results show that \system\ enables accurate recovery of continuous membership intervals for interval-level membership inference. 
Note that recovering temporal membership intervals is inherently more challenging than timestamp-level inference, as it requires not only identifying the correct timestamps but also accurately localizing interval boundaries and preserving the continuity of membership segments. Unlike binary classification, interval-level prediction is evaluated based on the overlap between predicted and ground-truth intervals, where even small boundary errors can substantially reduce the score. 
Although the absolute I-F1 values are lower than the timestamp-level results in some settings, particularly for vision models, \system\ consistently provides the most accurate interval recovery among all evaluated methods. These results demonstrate that \system\ substantially advances interval-level membership inference under this considerably more challenging prediction task.

\subsection{Robustness to Shadow Model Heterogeneity}\label{sc:Generalization Under Mismatched Shadow Models}

\begin{figure*}[!t]
    \centering
        \subfloat[Target-shadow datasets]{%
        \begin{minipage}[b]{0.245\textwidth}
        \includegraphics[width=\textwidth]{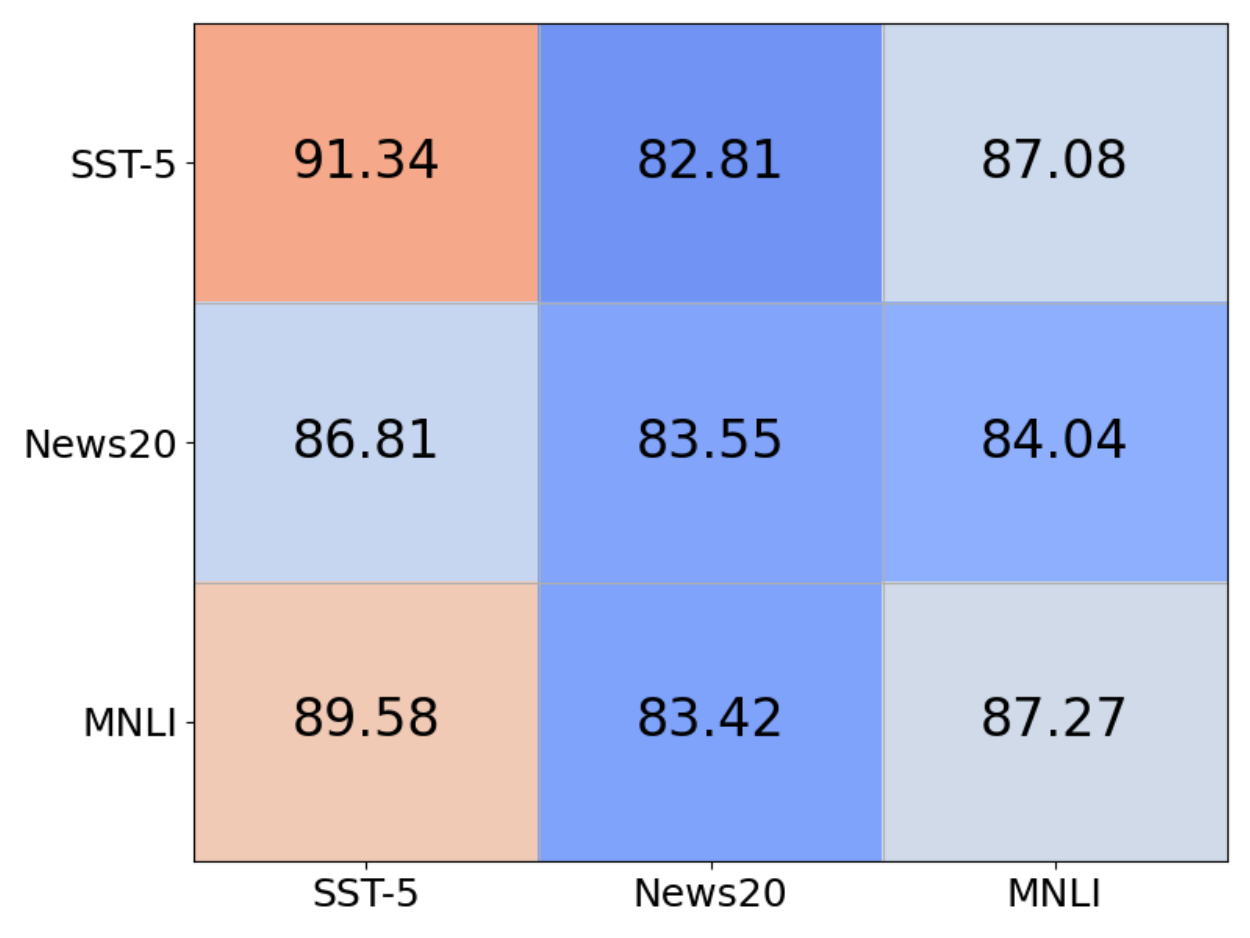} 

        \label{fig:plot_dataset_transfer_f1}
        \end{minipage}%
        }
    \subfloat[Target-shadow models]{%
    \begin{minipage}[b]{0.245\textwidth}
        \includegraphics[width=\textwidth]{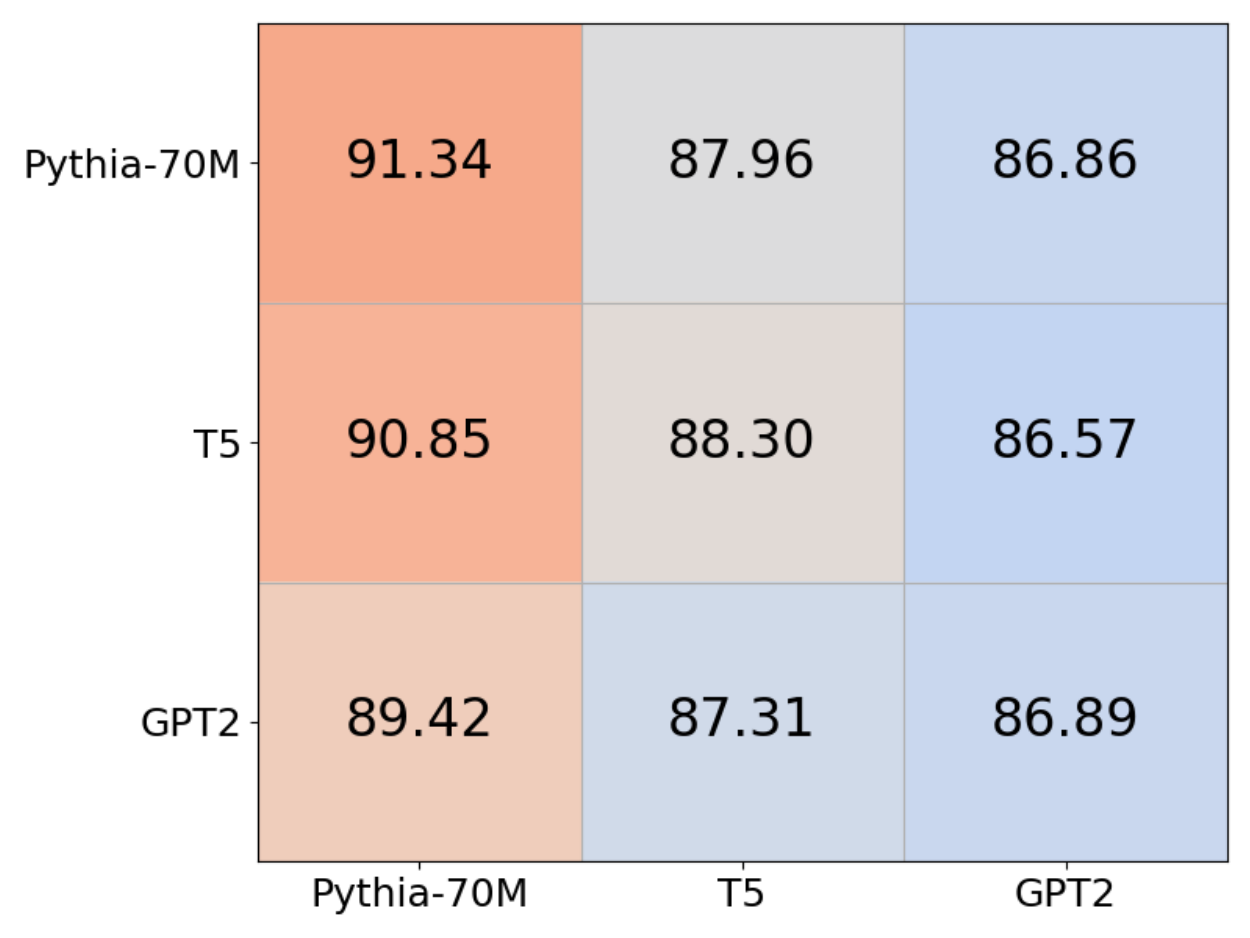} 

        \label{fig:plot_model_transfer_f1}
    \end{minipage}%
    }
     \subfloat[Target-shadow update methods]{%
     \begin{minipage}[b]{0.245\textwidth}
        \includegraphics[width=\textwidth]{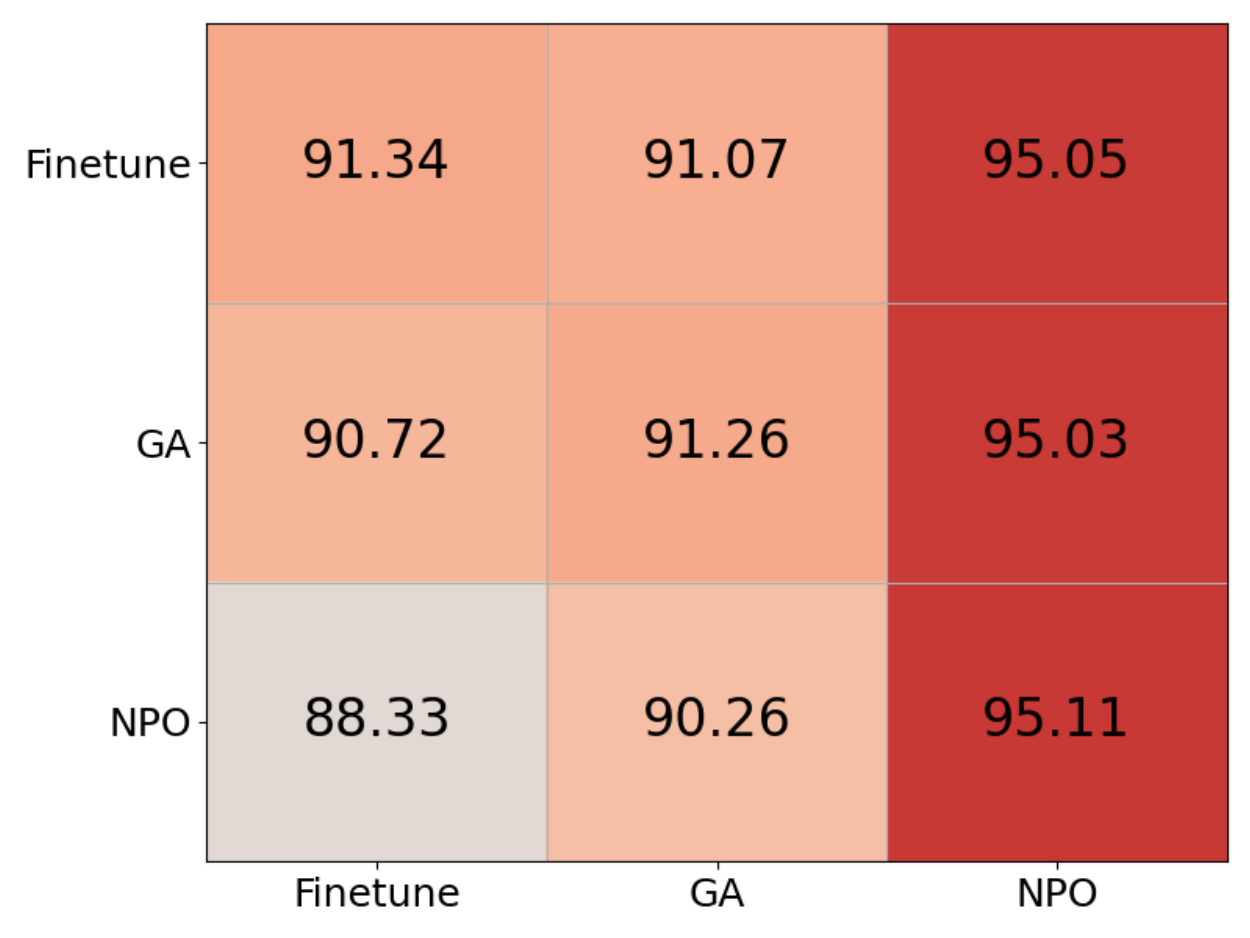}

        \label{fig:plot_unlearning_transfer_f1}
     \end{minipage}%
     }
    \subfloat[Target-shadow update volume]{%
     \begin{minipage}[b]{0.245\textwidth}
        \includegraphics[width=\textwidth]{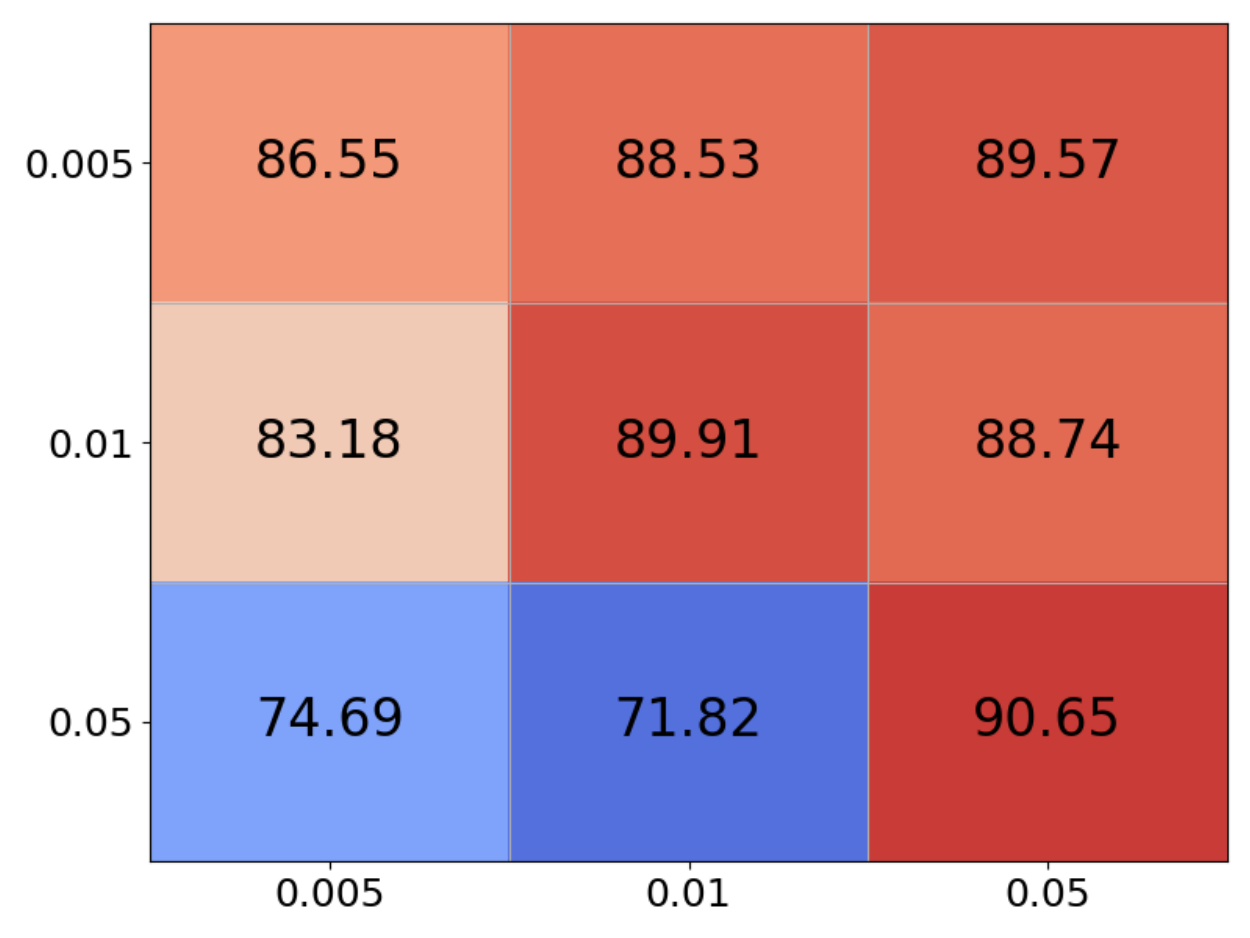}

        \label{fig:Target-shadow update size_T_F1}
     \end{minipage}%
     }

   \caption{Timestamp-level performance of \system\  when the shadow model mismatches the target model. The horizontal and vertical axes correspond to the target and shadow configurations, respectively, including the model, dataset and update method. }
  \label{figure:Attack-transferability-F1}
\end{figure*}

\begin{figure*}[!t]
    \centering
     \includegraphics[width=0.25\textwidth]{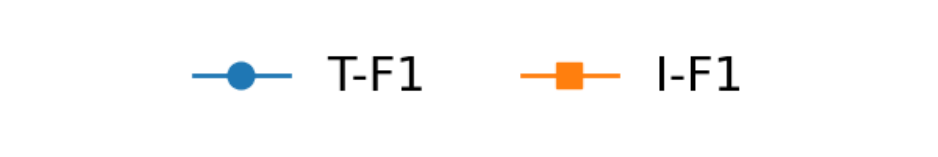}
     
    \subfloat[\small Sliding-window size]{%
    \begin{minipage}[b]{0.235\textwidth}
        \includegraphics[width=\textwidth]{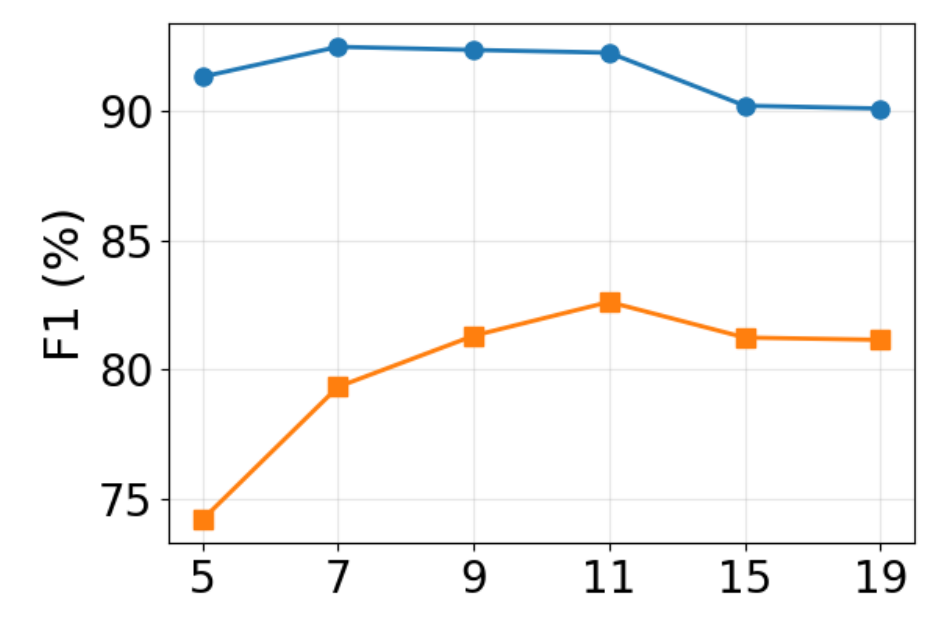} 

\label{fig:parameter_window_size}
    \end{minipage}%
    }
     \subfloat[\small Per-timestamp update volume]{%
     \begin{minipage}[b]{0.235\textwidth}
        \includegraphics[width=\textwidth]{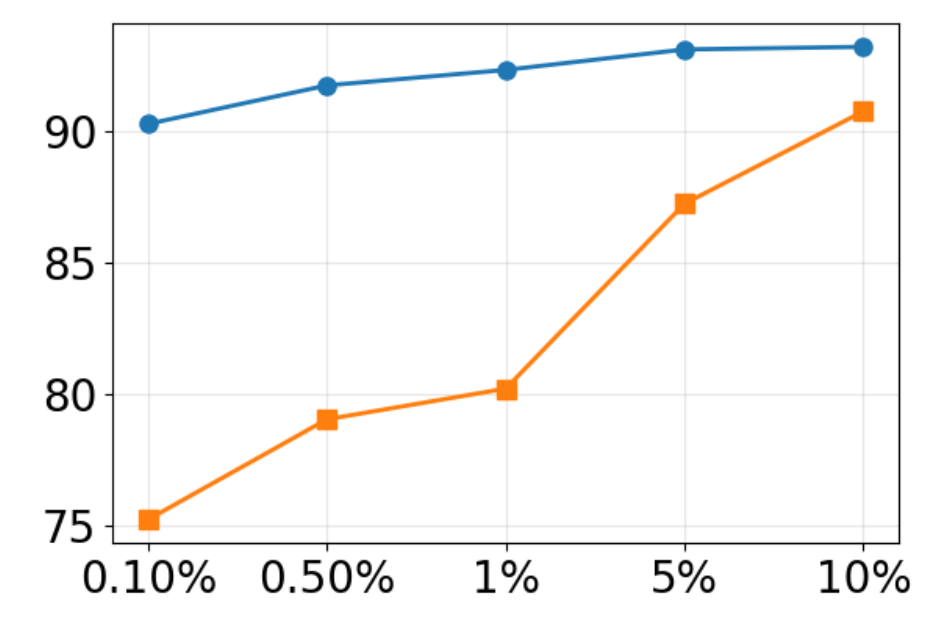} 

\label{fig:parameter_update_size}
     \end{minipage}%
     }
           \subfloat[\small Per-sample update count]{%
           \begin{minipage}[b]{0.235\textwidth}
        \includegraphics[width=\textwidth]{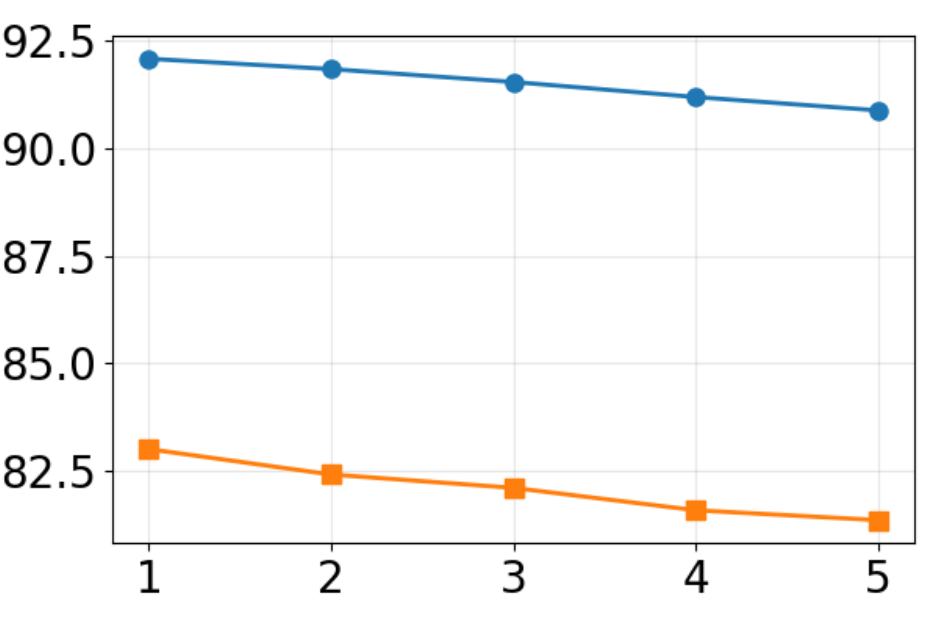} 

\label{fig:parameter_insert}
           \end{minipage}%
           }
         \subfloat[\small Per-sample mem. duration]{%
         \begin{minipage}[b]{0.235\textwidth}
        \includegraphics[width=\textwidth]{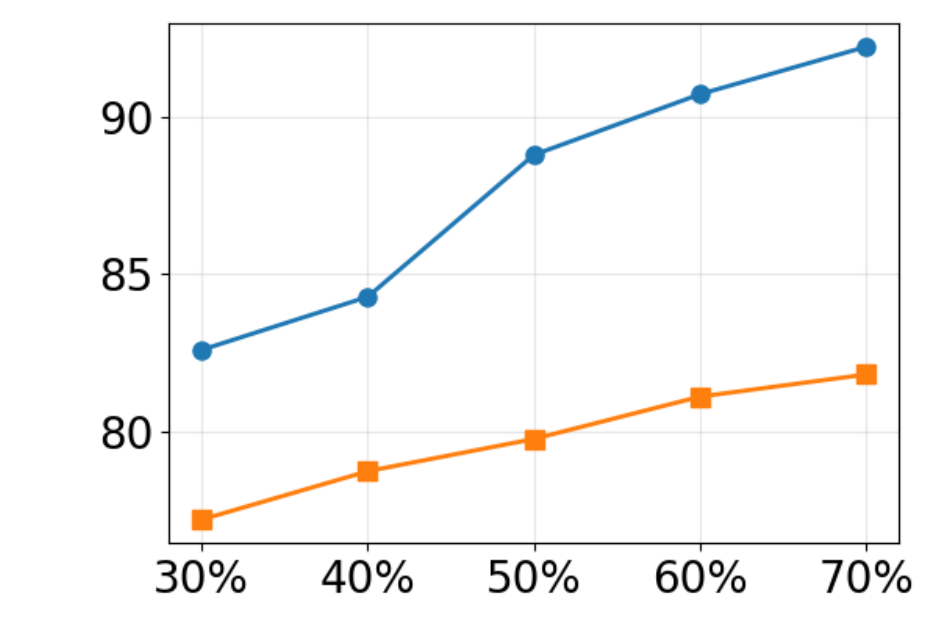} 

\label{fig:parameter_interval_member}
         \end{minipage}%
         }
     \caption{Performance of \system\ under various parameter configurations (Pythia-70M, SST-5)}  
    \label{figure:factor-analysis}
\end{figure*}

In practice, an adversary’s knowledge of the shadow dataset, model architecture, and update algorithm may not align with those used by the target model. In this section, we conduct a comprehensive set of experiments evaluating \system's performance across varying mismatched scenarios, including mismatched datasets, mismatched model architecture, and mismatched update algorithms. Due to space limit, we only report the T-F1 results for the three LMs in this section. The I-F1 performance can be found in Appendix~\ref{appendix:shadow-model-mismatch}. 

\noindent{\bf Mismatched Datasets}.
We fix the target model to Pythia-70M and vary the shadow and target datasets among SST5, News20, and MNLI. Figure~\ref{fig:plot_dataset_transfer_f1} reports the T-F1 of \system\ under these settings. As expected, \system\ achieves the highest T-F1 when the shadow and target datasets coincide. Notably, even when the shadow model is trained on a different dataset, \system\ remains highly effective, exhibiting a marginal reduction in T-F1 of at most 4.95\% across all settings.

\noindent{\bf Mismatched Model Architectures}.
We fix SST5 as the target dataset and vary the shadow and target models across three architectures: Pythia-70M, T5, and GPT-2. Figure~\ref{fig:plot_model_transfer_f1} reports the T-F1 results. In all cases, \system\ achieves its best performance when the shadow and target models share the same architecture. Nevertheless, \system\ remains robust when the two models use different architectures, with a T-F1 degradation of no more than 2.1\% relative to the best-performing shadow architecture.

\noindent{\bf Mismatched Update Algorithms}.
We use Pythia-70M for both the shadow and target models and fix SST5 as the training dataset for both models. We then vary the update algorithms used by the shadow and target models. As shown in Figure~\ref{fig:plot_unlearning_transfer_f1}, \system\ achieves the highest T-F1 when the same update algorithm is used for both models. More importantly, when the shadow and target models employ different update algorithms, \system\ experiences only marginal performance degradation, with T-F1 remaining at least 88.33\% across all settings. 

\noindent\textbf{Mismatched Update Volume.}
We fix the dataset to SST-5 and the target model to Pythia-70M, and vary the shadow and target per-timestamp update volumes among $0.5\%$, $1\%$, and $5\%$ of the training data size. Figures~\ref{fig:Target-shadow update size_T_F1} 
reports the corresponding T-F1 and I-F1 of \system, respectively. Both metrics peak when the update volumes match. When the shadow update volume is no larger than the target's, T-F1 and I-F1 decrease by at most $1.91\%$ and $4.04\%$, respectively. Larger drops occur when the shadow model uses $5\%$ of the training data per timestamp while the target uses $0.5\%$ or $1\%$, with T-F1 decreasing by up to $18.09\%$.  

In summary, these results indicate that \system\ captures fundamental behavioral changes induced by model updates, thus enables effective transfer across diverse datasets, model architectures, update strategies, and update volumes. 

\subsection{Parameter Sensitivity Analysis}\label{sec:Parameter-Sensitivity-Analysis}

\begin{table*}[!t]
\centering
\caption{Timestamp-level accuracy of \system\ and baselines for language models. The best results for each setting are marked in \gray{gray}.}
\begin{tabular}{c|c|ccccccccc}
\hline
\multirow{2}{*}{\textbf{Model}} & \multirow{2}{*}{\textbf{Dataset}}  & \multicolumn{3}{c|}{\textbf{TMIA (Ours)}}                                                                      & \multicolumn{3}{c|}{\textbf{Timestamp-wise MIA}}                                                               & \multicolumn{3}{c}{\textbf{Update-wise MIA}}                                              \\ \cline{3-11} 
                                &                                   & \multicolumn{1}{c|}{\textbf{T-Pre}} & \multicolumn{1}{c|}{\textbf{T-Rec}} & \multicolumn{1}{c|}{\textbf{T-F1}} & \multicolumn{1}{c|}{\textbf{T-Pre}} & \multicolumn{1}{c|}{\textbf{T-Rec}} & \multicolumn{1}{c|}{\textbf{T-F1}} & \multicolumn{1}{c|}{\textbf{T-Pre}} & \multicolumn{1}{c|}{\textbf{T-Rec}} & \textbf{T-F1} \\ \hline
\multirow{2}{*}{Llama-3B}       & Squad                             & \multicolumn{1}{c|}{99.48}          & \multicolumn{1}{c|}{81.59}          & \multicolumn{1}{c|}{\gray{89.65}}         & \multicolumn{1}{c|}{53.87}          & \multicolumn{1}{c|}{95.70}          & \multicolumn{1}{c|}{68.79}         & \multicolumn{1}{c|}{32.87}          & \multicolumn{1}{c|}{47.11}          & 38.72         \\
                                & Simpleqa                          & \multicolumn{1}{c|}{98.25}          & \multicolumn{1}{c|}{92.15}          & \multicolumn{1}{c|}{\gray{95.25}}         & \multicolumn{1}{c|}{59.05}          & \multicolumn{1}{c|}{99.98}          & \multicolumn{1}{c|}{74.06}         & \multicolumn{1}{c|}{57.83}          & \multicolumn{1}{c|}{47.56}          & 52.19         \\ \hline
\multirow{2}{*}{Llama-8B}       & Squad                             & \multicolumn{1}{c|}{96.22}          & \multicolumn{1}{c|}{85.16}          & \multicolumn{1}{c|}{\gray{90.35}}         & \multicolumn{1}{c|}{50.61}          & \multicolumn{1}{c|}{94.98}          & \multicolumn{1}{c|}{65.88}         & \multicolumn{1}{c|}{32.34}          & \multicolumn{1}{c|}{46.66}          & 38.21         \\
                                & Simpleqa                          & \multicolumn{1}{c|}{95.61}          & \multicolumn{1}{c|}{75.45}          & \multicolumn{1}{c|}{\gray{84.34}}         & \multicolumn{1}{c|}{55.08}          & \multicolumn{1}{c|}{99.95}          & \multicolumn{1}{c|}{70.88}         & \multicolumn{1}{c|}{56.73}          & \multicolumn{1}{c|}{45.66}          & 50.6          \\ \hline
\end{tabular}
\label{tab:LLM-TL}  
\end{table*}

In this section, we evaluate the effect of four key parameters on \system's performance: (1) \textit{BiLSTM sliding-window size}; (2)  \textit{per-timestamp update volume} (total number of inserted and deleted samples per timestamp), (3) \textit{per-sample update count}, and (4) \textit{per-sample membership duration} (how long a sample participates in training). 
Figure \ref{figure:factor-analysis} reports the results for the Pythia-70M model trained on SST-5 dataset. Additional results for other settings are in Appendix~\ref{Appendix:Parameter-Sensitivity-Analysis}. 

\textbf{Sliding-window Size.}
Figure~\ref{fig:parameter_window_size} reports the performance of \system\ under different sliding-window sizes in the BiLSTM model. We observe that attack performance improves consistently as the sliding window size increases. Notably, a window size of 7 yields the best T-F1 performance, while a window size of 11 achieves the highest I-F1 performance. Further increasing the window size provides no additional benefit and instead slightly degrades both T-F1 and I-F1 scores. These results suggest that there exists an optimal window size for balancing timestamp- and interval-level inference, beyond which larger windows offer diminishing benefits. 

\textbf{Per-timestamp Update Volume.} 
We vary the update volume, which is measured as the ratio to the training data size, from 0.1\% to 10\%, and evaluate \system’s performance. As shown in Figure~\ref{fig:parameter_update_size}, as the update volume increases, both T-F1 and I-F1 values increase accordingly. This is expected as higher update volumes induce stronger perturbations to the model parameters, which enables the update effects to persist over longer periods, leading samples to form more stable temporal intervals to be identified. 

\textbf{Per-sample Update Count.}
Figure~\ref{fig:parameter_insert} shows \system’s performance as the per-sample update count rises from 1 to 5. Both T-F1 and I-F1 decline slightly, likely due to more frequent membership changes fragmenting temporal patterns and complicating entry/exit identification. The decrease is minimal—only 1.3\% in T-F1 and 1.98\% in I-F1. The limited performance drop indicates that \system\ is robust to large per-sample update counts.

\textbf{Per-sample Membership Duration.} 
Figure~\ref{fig:parameter_interval_member} presents \system’s performance under varying membership durations, measured as the ratio to the total number of timestamps (80 in total). \system’s performance consistently improves as the length of the membership duration increases across all datasets. This is because longer membership intervals allow confidence score changes to stabilize over time, making entry and exit points easier to identify. These results show that longer participation in training amplifies temporal membership leakage and poses greater privacy risks.

\subsection{Extending to Large Generative Models}
\label{sc:llm}

To assess \system’s effectiveness on large generative language models, we experiment with the LLaMA~3 family~\cite{dubey2024llama}, a widely used open-weight LLM. We evaluate two models with 3B and 8B parameters, denoted LLaMA-3B and LLaMA-8B. Two question-answering datasets are used: \textit{SQuAD}~\cite{rajpurkar2016squad} ($\approx$100K QA pairs) and \textit{SimpleQA}~\cite{wei2024measuring} ($\approx$4K QA pairs). We simulate continual updates across 40 timestamps, with each operation affecting 1\% of the training set. Updates are applied using LoRA-based fine-tuning~\cite{liu2025lune}.

Since prediction confidence scores are not directly available for generative language models, we instead rely on the model’s \textit{perplexity} (PPL). Perplexity is a standard metric for evaluating a language model’s ability to predict the next token in a sequence \cite{jelinek1977perplexity}, with lower values indicating better predictive performance and higher confidence. We adapt \system\ to the generative language model setting by using the sequence of the PPL values across all model snapshots as the input.
For the baseline methods, we first implement a timestamp-wise MIA (TS-MIA) by applying a PPL-based membership inference attack \cite{carlini2021extracting} independently to each model snapshot. For the update-wise MIA (UP-MIA), we modify the delta attack \cite{jagielski2023combine} by replacing confidence scores with PPL values.

Table~\ref{tab:LLM-TL} reports the timestamp-level attack performance of \system\ and two baseline methods. \system\ consistently achieves very high timestamp-level precision (T-Pre), ranging from 95.61\% to 99.48\%, together with strong recall (T-Rec), resulting in T-F1 scores above 84\% across all settings. In contrast, the timestamp-wise MIA attains high recall but suffers from low precision, while the update-wise MIA performs poorly on both precision and recall, leading to substantially lower T-F1 scores. The results of interval-level attack performance are similar and can be found in Appendix~\ref{appendix:I-F1-LLM}.

\section{Robustness Under Realistic Updates}
\label{sc:robustness}

To further assess the practicality and robustness of \system, we evaluate its performance under three realistic scenarios: (1)  Model updates exhibit temporal correlations; 
(2) Model outputs are modified with reduced precision; 
and 
(3) The adversary observes only a subset of model snapshots. 
These evaluations are designed to examine whether \system\ can generalize to realistic conditions in which the attacker has incomplete observations and the target model differs from the original full-precision setting.

\subsection{Attacking Correlated Model Updates}
\label{subsec:correlated}

\begin{table}[t!]
\centering
\caption{Performance of \system\ under correlated model updates (SST-5, Pythia-70M).}
\label{tab:correlated}
\resizebox{\linewidth}{!}{%
\begin{tabular}{lcccccc}
\toprule
& \multicolumn{3}{c}{\textbf{Timestamp-level}} & \multicolumn{3}{c}{\textbf{Interval-level}} \\
\cmidrule(lr){2-4}\cmidrule(lr){5-7}
\textbf{Updates} & \textbf{T-Pre} & \textbf{T-Rec} & \textbf{T-F1} & \textbf{I-Pre} & \textbf{I-Rec} & \textbf{I-F1} \\
\midrule
Uncorrelated & 88.81 & 91.51 & 90.14 & 72.10 & 89.92 & 80.03 \\
Correlated insertion     & 88.50 & 91.70 & 90.07 & 71.50 & 90.60 & 79.92 \\
Correlated removal    & 88.30 & 91.20 & 89.73 & 71.40 & 89.80 & 79.55 \\
Correlated (both)      & 87.60 & 91.80 & 89.65 & 70.20 & 91.30 & 79.37 \\
\bottomrule
\end{tabular}}
\end{table}

So far, our evaluation has focused on uncorrelated model updates, where the samples involved in each update are drawn independently. In practice, however, ML systems often receive correlated updates, such as data that share the same topic, belong to the same class, or follow the same distribution. Such correlations may alter the influence of individual updates and, consequently, affect the effectiveness of \system. 

To investigate this effect, we use the Pythia-70M model trained on the SST-5 dataset as the victim model. We simulate correlated updates by constructing each model update from samples (sentences) belonging to a single sentiment class. Since all samples in an update share the same label, they are semantically correlated. In contrast, samples drawn from different sentiment classes are treated as uncorrelated.

We evaluate \system\ under the following four update settings: 
(1) {\bf Correlated insertion}: insert correlated samples while removing uncorrelated samples;
(2) {\bf Correlated removal}: remove correlated samples while inserting uncorrelated samples;
(3) {\bf Correlated insertion \& removal}: both inserted and removed samples are correlated; and 
(4) {\bf Uncorrelated updates}: both inserted and removed samples are uncorrelated.
To reflect a realistic attack scenario, we assume the adversary has no prior knowledge of whether the target model is updated with correlated or uncorrelated data. Consequently, the attacker trains shadow models using only uncorrelated updates.

Table~\ref{tab:correlated} shows that \system\ remains highly effective even when the victim model undergoes correlated updates, achieving attack accuracy comparable to the uncorrelated baseline. Although correlated updates lead to a slight reduction in performance, the degradation is modest: T-F1 and I-F1 change by at most 0.49\% and 0.66\%, respectively. 
The slight performance drop likely because correlated samples more similar model updates, reducing the diversity of update signatures and making insertion and removal events somewhat less distinguishable. However, they still leave consistent traces in the model parameters which allow  \system\ to generalize effectively even though its shadow models are trained exclusively on uncorrelated updates. 

\subsection{Attacking Model Outputs with Reduced-Precision}
\label{subsec:quantization}

\begin{table}[t!]
\centering
\caption{Performance of \system\ against reduced-precision model outputs by quantization 
(Pythia-70M model).}
\label{tab:quantization}
\setlength{\tabcolsep}{4.5pt}
\begin{tabular}{lcccccc}
\toprule
& \multicolumn{3}{c}{\textbf{Timestamp-level}} & \multicolumn{3}{c}{\textbf{Interval-level}} \\
\cmidrule(lr){2-4}\cmidrule(lr){5-7}
\textbf{Setting} & \textbf{T-Pre} & \textbf{T-Rec} & \textbf{T-F1} & \textbf{I-Pre} & \textbf{I-Rec} & \textbf{I-F1} \\
\midrule
Full precision & 88.81 & 91.51 & 90.14 & 72.10 & 89.92 & 80.03 \\
8-bit          & 83.64 & 77.26 & 80.32 & 70.61 & 74.53 & 72.52 \\
4-bit          & 75.72 & 54.33 & 63.27 & 62.52 & 62.49 & 62.50 \\
\bottomrule
\end{tabular}
\end{table}

\begin{table}[t!]
\centering
\caption{Performance of \system\ under partial access to model updates (Pythia-70M model).}
\label{tab:access}
\setlength{\tabcolsep}{4.5pt}
\begin{tabular}{ccccccc}
\toprule
\textbf{Access to}& \multicolumn{3}{c}{\textbf{Timestamp-level}} & \multicolumn{3}{c}{\textbf{Interval-level}} \\
\cmidrule(lr){2-4}\cmidrule(lr){5-7}
\textbf{model updates} & \textbf{T-Pre} & \textbf{T-Rec} & \textbf{T-F1} & \textbf{I-Pre} & \textbf{I-Rec} & \textbf{I-F1} \\
\midrule
Full sequence& 88.81 & 91.51 & 90.14 & 72.10 & 89.92 & 80.03 \\
$50\%$ of snapshots & 69.07 & 76.98 & 72.82 & 60.54 & 69.39 & 64.66 \\
$25\%$ of snapshots & 40.42 & 60.55 & 48.48 & 37.96 & 35.10 & 36.47 \\
\bottomrule
\end{tabular}
\end{table}

\begin{table*}[h!]
\center
\caption{Performance of three defense mechanisms against \system.}
\label{tab:potential mitigation}
\begin{tabular}{c|c|ccc|ccc|cc}
\hline
\multirow{2}{*}{\textbf{Dataset/Model}} & \multirow{2}{*}{\textbf{Defense}} & \multicolumn{3}{c|}{\textbf{Timestamp-level}}                                             & \multicolumn{3}{c|}{\textbf{Interval-level}}                                              & \multicolumn{2}{c}{\textbf{Model Utility}}                        \\ \cline{3-10} 
                                        &                                   & \multicolumn{1}{c|}{\textbf{T-Pre}} & \multicolumn{1}{c|}{\textbf{T-Rec}} & \textbf{T-F1} & \multicolumn{1}{c|}{\textbf{I-Pre}} & \multicolumn{1}{c|}{\textbf{I-Rec}} & \textbf{I-F1} & \multicolumn{1}{c|}{\textbf{Train. Acc}} & \textbf{Test. Acc} \\ \hline
\multirow{6}{*}{SST-5/Pythia-70M}      & W/o defense                       & \multicolumn{1}{c|}{88.78}          & \multicolumn{1}{c|}{94.06}          & 91.34         & \multicolumn{1}{c|}{80.89}          & \multicolumn{1}{c|}{87.70}          & 84.16         & \multicolumn{1}{c|}{99.18}                 & 43.26                \\ \cline{2-10} 
                                        & Label-only output                 & \multicolumn{1}{c|}{84.79}          & \multicolumn{1}{c|}{91.44}          & 87.99         & \multicolumn{1}{c|}{67.90}          & \multicolumn{1}{c|}{83.39}          & 74.85         & \multicolumn{1}{c|}{99.18}                 & 43.25                \\ \cline{2-10} 
                                        & Dropout$_{p=30\%}$                 & \multicolumn{1}{c|}{80.71}          & \multicolumn{1}{c|}{71.77}          & 75.36         & \multicolumn{1}{c|}{73.63}          & \multicolumn{1}{c|}{72.84}          & 73.22         & \multicolumn{1}{c|}{65.12}                 & 40.13                \\ \cline{2-10} 
                                        & Dropout$_{p=50\%}$                  & \multicolumn{1}{c|}{76.39}          & \multicolumn{1}{c|}{46.99}          & 58.18         & \multicolumn{1}{c|}{64.15}          & \multicolumn{1}{c|}{43.51}          & 51.82         & \multicolumn{1}{c|}{60.11}                 & 41.24                \\ \cline{2-10} 
                                        & DP$_{\epsilon_{total}=30}$                          & \multicolumn{1}{c|}{71.60}          & \multicolumn{1}{c|}{24.03}          & 35.99         & \multicolumn{1}{c|}{27.47}          & \multicolumn{1}{c|}{14.93}          & 19.35         & \multicolumn{1}{c|}{54.67}                 & 40.11                \\ \cline{2-10} 
                                        & DP$_{\epsilon_{total}=10}$                        & \multicolumn{1}{c|}{63.35}          & \multicolumn{1}{c|}{10.12}          & 17.47         & \multicolumn{1}{c|}{6.62}           & \multicolumn{1}{c|}{4.28}           & 5.20          & \multicolumn{1}{c|}{46.32}                 & 38.43                \\ \hline
\multirow{6}{*}{CIFAR10/SimpleCNN}     & W/o defense                       & \multicolumn{1}{c|}{80.22}          & \multicolumn{1}{c|}{76.39}          & 78.26         & \multicolumn{1}{c|}{79.05}          & \multicolumn{1}{c|}{53.26}          & 63.64         & \multicolumn{1}{c|}{100.00}                & 72.12                \\ \cline{2-10} 
                                        & Label-only output                 & \multicolumn{1}{c|}{75.95}          & \multicolumn{1}{c|}{75.61}          & 75.78         & \multicolumn{1}{c|}{65.49}          & \multicolumn{1}{c|}{51.37}          & 57.56         & \multicolumn{1}{c|}{100.00}                & 72.12                \\ \cline{2-10} 
                                        & Dropout$_{p=30\%}$                   & \multicolumn{1}{c|}{75.01}          & \multicolumn{1}{c|}{73.33}          & 73.53         & \multicolumn{1}{c|}{62.88}          & \multicolumn{1}{c|}{50.52}          & 56.03         & \multicolumn{1}{c|}{98.12}                 & 71.23                \\ \cline{2-10} 
                                        & Dropout$_{p=50\%}$                    & \multicolumn{1}{c|}{70.62}          & \multicolumn{1}{c|}{54.38}          & 61.44         & \multicolumn{1}{c|}{40.53}          & \multicolumn{1}{c|}{18.83}          & 25.76         & \multicolumn{1}{c|}{86.22}                 & 71.51                \\ \cline{2-10} 
                                        & DP$_{\epsilon_{total}=30}$                        & \multicolumn{1}{c|}{69.58}          & \multicolumn{1}{c|}{27.62}          & 39.54         & \multicolumn{1}{c|}{34.43}          & \multicolumn{1}{c|}{17.88}          & 23.54         & \multicolumn{1}{c|}{56.65}                 & 52.33                \\ \cline{2-10} 
                                        & DP$_{\epsilon_{total}=10}$                        & \multicolumn{1}{c|}{67.13}          & \multicolumn{1}{c|}{16.20}          & 26.10         & \multicolumn{1}{c|}{27.08}          & \multicolumn{1}{c|}{10.10}          & 14.77         & \multicolumn{1}{c|}{45.28}                 & 41.31                \\ \hline
\end{tabular}
\end{table*}

In practice, an adversary may only have access to reduced-precision model outputs rather than full-precision predictions. Such outputs commonly arise because deployed machine learning systems models often undergo post-processing, such as quantization, pruning, distillation, or output calibration. These techniques can alter or smooth the model outputs. Since \system\ relies on temporal variations in model outputs, such reduced-precision outputs may weaken the temporal membership signals exploited by the attack. 

To evaluate whether \system\ remains effective against model outputs with reduced precision, we configure the target model using per-channel symmetric uniform quantization~\cite{jacob2018quantization} on the Pythia-70M model. The weights in each linear layer are quantized to $b$ bits, while activations remain in full precision. We consider two quantization levels, $b\in\{4,8\}$. 
We assume a strong adversary who knows that the target model has been quantized and is aware of the quantization algorithm and configuration. Accordingly, the adversary applies the same quantization procedure to its shadow models. Such an assumption is reasonable in practice, as many deployed models publicly disclose their quantization method and bit width. 

Table~\ref{tab:quantization} reports the performance of \system\ on both full-precision and quantized models. As expected, reducing the output precision degrades the attack performance, with T-F1 and I-F1 decreasing by 26.37\% and 17.53\% for 4-bit quantization, as well as 9.82\% and 7.51\% for 8-bit quantization, respectively, compared with the full-precision model. The degradation occurs because quantization suppresses small variations in the model outputs, making temporal membership signals less distinguishable across consecutive model snapshots.
Nevertheless, \system\ remains effective even under aggressive 4-bit quantization, achieving T-F1 and I-F1 scores of 63.3\% and 62.5\%, respectively. This result indicates that  quantization does not eliminate the changes in model updates, allowing \system\ to reliably infer temporal membership.

\subsection{Attacking with Partial Observation of Model Updates}
\label{subsec:access}

Our evaluation so far has assumed that the adversary can query every model snapshot throughout the model update sequence. In practice, however, an adversary may have only intermittent or limited access to the deployed model and thus observe only a subset of the snapshots.

To evaluate the robustness of \system\ under such restricted access, we consider two scenarios where the adversary observes only 25\% and 50\% of the complete sequence of model snapshots, respectively. For each ratio, we randomly sample the accessible snapshots from the entire update sequence while preserving their original temporal order. This simulates realistic scenarios where an adversary can periodically query the target model but cannot continuously monitor every update.

As shown in Table~\ref{tab:access}, reducing snapshot availability degrades the performance of \system. This degradation is expected because some insertion/removal events may occur between two observed snapshots and therefore remain undetected, reducing both attack precision and recall. Nevertheless, \system\ remains effective even with limited observations. When the adversary has access to only 50\% of the model snapshots, \system\ still achieves a T-F1 score of 72.82\% and an I-F1 score of 64.66\%, demonstrating that periodic access to model snapshots still  poses a substantial privacy threat.

\section{Potential Defense}

Having established the effectiveness of \system, we next evaluate the effectiveness of potential defense mechanisms against the attack. 

We adapt three defense strategies: \textit{label-only output}~\cite{chen2021machine, choquette2021label}, \textit{dropout}~\cite{kaya2020effectiveness,yin2021defending}, and \textit{differential privacy} (DP)~\cite{dwork2006differential}, which have been shown effective against MIAs in static models, to our setting:  

\begin{ditemize}
\item \textbf{Label-only output}: each model snapshot returns only its predicted class label, without exposing confidence scores. To construct \system's attack features, we assign a binary correctness indicator of 1 if the predicted label matches, and 0 otherwise. Then we replace the confidence-score trajectory with the binary indicators to construct the attack features. 
\item \textbf{Dropout}: dropout is enabled during both model training (including subsequent updates) and inference. A fraction $p$ of units in the input and fully connected layers are randomly deactivated, so each snapshot's outputs are also subject to dropout at query time. Following prior work \cite{kaya2020effectiveness,yin2021defending}, We use dropout rates $p\in\{30\%,50\%\}$. 
\item {\bf Differential privacy}: we deploy \textit{Differentially Private Stochastic Gradient Descent (DP-SGD)}~\cite{abadi2016deep}, a state-of-the-art DP mechanism, to each model snapshot. 
The total privacy loss is obtained by summing the RDP guarantees across all timestamps, leveraging the composability property of RDP~\cite{mironov2017renyi}. Finally, the accumulated RDP guarantee is converted into the corresponding $(\varepsilon, \delta)$-DP guarantee using the standard RDP-to-DP conversion~\cite{mironov2017renyi}. 
We use the privacy budget $\epsilon\in\{10,30\}$ and fix $\delta=5\times 10^{-4}$ in our evaluation. 
\end{ditemize}

Table~\ref{tab:potential mitigation} reports the performance of the three defense mechanisms. All three consistently reduce attack performance at both the timestamp and interval levels, with DP offering the strongest protection and achieving the largest reductions in both T-F1 and I-F1.
On the other hand, the label-only output defense yields the weakest mitigation power, resulting in modest reductions in both T-F1 and I-F1 (3\% and 11\% reduction, respectively). 

Table \ref{tab:potential mitigation} also reveals a privacy-utility trade-off. While DP offers the strongest protection, it also incurs the most severe loss in model utility.
In contrast, the label-only output defense largely preserves model utility but provides only marginal privacy protection against \system. These results suggest that dropout offers the most favorable balance between privacy protection and model utility.

\section{Conclusion}

In this paper, we first studied the problem about temporal membership inference in continually updated deep learning models and formalize \textit{timestamp-level} and \textit{interval-level} membership inference. We proposed \system, a novel temporal attack that exploits membership dynamics to infer whether and when specific samples are involved in the training process of the target model. Extensive experiments demonstrated the effectiveness of \system. 

This work opens several promising directions for future research. First, while we focus on black-box attacks, extending to the white-box setting and investigating how model internals, such as gradients, interact with different update strategies remains an open question. Second, although \system\ was tested against common defenses, designing methods specifically to mitigate temporal membership leakage is largely unexplored. Finally, temporal membership inference raises fundamental questions about defining and auditing privacy for evolving models. Developing formal guarantees and benchmarks that account for temporal dynamics is both crucial and urgent.

\appendix
\section{Proof of Theorem~\ref{theorm:validinterval}}
\label{appendix:proof-theorem}

\begin{proof}
First, we have the following lemma that presents an important property of the representative timestamps output by Step 2. 

\begin{lemma}[Alternating Labels of Consecutive Representative Timestamps]\label{lemma1}
Let $\mathbf{\tau} = \{\tau_1, \ldots, \tau_K\}$ denote the set of representative timestamps. Then no two consecutive timestamps in $\tau$ share the same label; that is, for all $i = 1, \dots, K-1$, we have
\[
\ell_{\tau_i} \neq \ell_{\tau_{i+1}}.
\]

\end{lemma}

 \begin{proof} The proof is straightforward. If $\ell_{\tau_i} = \ell_{\tau_{i+1}}$, Step 1 would merge the groups of these two representative timestamps into a single larger group. The representative timestamp of the merged group would then be the one with the higher confidence score among $\ell_{\tau_i}$ and $\ell_{\tau_{i+1}}$.  
\end{proof}

Let $\mathbf{\tau} = \{\tau_1, \ldots, \tau_K\}$ denote the set of representative timestamps. Let $L = (\ell_{\tau_1}, \dots, \ell_{\tau_K})$ be the labels associated with each representative timestamps. Let $k_{E}$ and $k_{X}$ be the number of label 1 (Entry-point) and 2 (Exit-point) in $L$. Following Lemma~\ref{lemma1}, no consecutive labels in $L$ are the same, leading to the following three cases. For clarity of presentation, we use “E” and “X” to denote labels 1 and 2, respectively, in the following discussions.

\paragraph{Case 1: $k_{E} = k_{X}$.}  
Since no consecutive labels in $L$ are the same, the label sequence of $L$ must take one of the following two alternating forms:
\[
\underbrace{E, X, E, X, \ldots, E, X}_{2 k_E \text{ labels}}
\quad \text{or} \quad
\underbrace{X, E, X, E, \ldots, X, E}_{2 k_E \text{ labels}}.
\]

In the first form, Step 3 generates $k_E$ membership intervals, each corresponding to an adjacent $[E, X]$ pair.  
In the second form, Step 3 generates $k_E + 1$ membership intervals: the first interval is $[t_0, \tau_{X}]$, the last interval is $[\tau_{E}, t_N]$, and the remaining intervals correspond to the intermediate $[E, X]$ pairs, where $\tau_{X}$ and $\tau_{E}$ denote the timestamp of the first X and the last E label in $L$.

\paragraph{Case 2: $k_{E} = k_{X}+1$.}
For this case, the label sequence $L$ must take the following   form:
\[
\underbrace{E, X, E, X, \ldots, E, X, E}_{ k_E+k_X \text{ labels}}
\]
It is possible that $k_X=0$; then $L=\{E\}$. 
Then Step 3 constructs $k_X+1$ intervals: the first $k_X$ intervals correspond to the intermediate $[E, X]$ pairs, while the last interval is $[\tau_{E}, t_N]$, where $\tau_{E}$ denotes the timestamp of the last E label in $L$. When $L=\{E\}$, it only constructs one interval $[\tau_{E}, t_N]$. 

\paragraph{Case 3: $k_{X} = k_{E}+1$.}
For this case, the label sequence $L$ must take the following   form:
\[
\underbrace{X, E, X, E, \ldots, E, X}_{ k_E+k_X \text{ labels}}
\]
It is possible that $k_E=0$; then $L=\{X\}$. 
Then Step 3 constructs $k_E+1$ intervals: the first interval is $[t_0, \tau_{X}]$, where $\tau_{X}$ denotes the timestamp of the first X label in $L$, while the remaining $k_E$ intervals correspond to the intermediate $[E, X]$ pairs. When $L=\{X\}$, it only constructs one interval $[t_0, \tau_{X}]$. 

In all three cases, the constructed intervals satisfy the
three requirements in
Definition~\ref{def:valid-intervals}. First, every
entry point precedes its matched exit point because the
representative timestamps are processed in chronological
order. Second, every explicit exit point is matched with a
unique preceding entry point: it is either paired with the
immediately preceding entry representative or, when it is the
first representative, with the implicit entry point $t_0$.
Moreover, because the labels strictly alternate, only the
final entry representative can remain without an explicit
exit point. Third, for any two consecutively constructed
intervals, the exit index of the first interval strictly
precedes the entry index of the second interval.  Thus the theorem follows. 
\end{proof}

\section{Experimental Setup}
\label{appendix:setup}

\subsection{Datasets}

We evaluate three text and three image datasets; Table~\ref{tab:dataset_stats} summarizes their statistics. SST-5~\cite{socher2013recursive} contains movie-review sentences for sentiment classification, News20~\cite{albishre2015effective} contains documents from 20 topics, and MNLI~\cite{wang2018glue} contains premise--hypothesis pairs for natural language inference. CIFAR-10~\cite{krizhevsky2009learning} and CINIC-10~\cite{darlow2018cinic} contain $32\times32$ color images from 10 classes, with CINIC-10 combining CIFAR-10 and ImageNet images. SVHN~\cite{netzer2011reading} contains cropped street-view house-number digits.

\begin{table}[t!]
\footnotesize
\renewcommand{\arraystretch}{0.9}
\centering
\caption{Statistics of datasets}
\label{tab:dataset_stats}
\begin{tabular}{c|c|c|c|c|c}
\hline
\textbf{Dataset} &
\textbf{Modality} &
\textbf{\begin{tabular}[c]{@{}c@{}}Feature\\ dim.\end{tabular}} &
\textbf{\# labels} &
\textbf{\begin{tabular}[c]{@{}c@{}}\# train\\ samples\end{tabular}} &
\textbf{\begin{tabular}[c]{@{}c@{}}\# test\\ samples\end{tabular}} \\
\hline
SST-5    & Text  & 512  & 5  & 8,544   & 2,210 \\
\hline
News20   & Text  & 512  & 20 & 11,314  & 7,532 \\
\hline
MNLI     & Text  & 512  & 3  & 50,000  & 9,815 \\
\hline
CIFAR-10 & Image & 3072 & 10 & 50,000  & 10,000 \\
\hline
SVHN     & Image & 3072 & 10 & 73,257  & 26,032 \\
\hline
CINIC-10 & Image & 3072 & 10 & 90,000  & 90,000 \\
\hline
\end{tabular}
\end{table}

\subsection{Models}
\begin{table}[t!]
\footnotesize
\renewcommand{\arraystretch}{0.95}
\centering
\caption{Architectural details of models used in the experiments}
\label{tab:model_arch}
\begin{tabular}{c|c|c|c}
\hline
\textbf{Model} &
\textbf{\# Layers} &
\textbf{Layer width} &
\textbf{\# of Parameters} \\
\hline
Pythia-70M   & 6  & 512  & 70M \\
\hline
GPT-2 Small  & 12 & 768  & 124M \\
\hline
T5           & 12 & 512  & 220M \\
\hline
ResNet-18    & 18 & \{64,128,256,512\} & 11.7M \\
\hline
SimpleCNN    & 4  & \{32,64,128,128\} & 1.2M \\
\hline
MobileNet    & 28 & \{32--1024\} & 4.2M \\
\hline
\end{tabular}
\end{table}

Table~\ref{tab:model_arch} summarizes the model architectures. Pythia-70M~\cite{biderman2023pythia} and GPT-2 Small~\cite{radford2019language} are autoregressive transformers; T5~\cite{raffel2020exploring} is a text-to-text transformer with ReLU activation, while GPT-2 uses GELU. The vision models are ResNet-18~\cite{he2016deep}, a residual CNN; SimpleCNN~\cite{chen2021machine}, with four ReLU convolutional layers, max-pooling, and two fully connected layers; and MobileNet~\cite{howard2017mobilenets}, an efficient CNN with ReLU activation.

\subsection{Data Updates}
\label{appendix:data-updates}
At each update, insertion and removal sets are independently sampled from the current non-member and member sets, respectively, allowing removed samples to be reinserted later. Each set contains 1\% of the current training set for SST-5, News20, CIFAR-10, and SVHN, 0.5\% for MNLI, and 0.1\% for CINIC-10. Updates operate at the sentence level for SST-5 and MNLI, document level for News20, and image level for image datasets. Repeated membership refers to at least two disjoint membership intervals. Among evaluation samples, approximately 1\% remain members throughout the observed sequence, 99\% undergo membership transitions, and none remain non-members throughout. 

\subsection{Setup of Attack Model}

The attack model uses a BiLSTM with feature dimension 32, followed by two fully connected layers: a 32-neuron ReLU hidden layer and a three-neuron softmax output layer. We randomly select 3,000 shadow samples for training and 1,200 target samples for evaluation.

\subsection{Update Methods}
\label{appendix:unlearning-methods}

We consider the following update methods in the paper: 
\begin{itemize}
\item \textit{Fine-tuning (FT)}: 
For removal, we fine-tune on forget samples with intentionally incorrect labels; for insertion, we use newly added samples with correct labels. 
\item \textit{Gradient Ascent (GA)}: 
GA~\cite{jang2023knowledge} increases the cross-entropy loss on forget samples, combined with a retain-set loss to preserve utility. 
\item \textit{Negative Preference Optimization (NPO)}:
NPO~\cite{zhang2024negative} suppresses forget-sample likelihood relative to a fixed reference model. We use $\beta=0.1$ and a retain-set loss. 
\item \textit{SCRUB}: SCRUB~\cite{kurmanji2023towards} trains a student model to diverge from a frozen teacher model on forget samples while preserving predictions and accuracy on retained samples. We disable checkpoint rewinding. The default batch size is 256, with learning rates of $10^{-3}$ for image classifiers and $10^{-5}$ for language models. 
\end{itemize}

\section{Additional Results: Empirical Characterization of Temporal Membership Signals}
\label{appendix:emp-charactirization}

\begin{figure}[!t]
	\centering
    \includegraphics[width=0.4\textwidth]{figures/legend-motivation.pdf}
    
     \includegraphics[width=0.38\textwidth]{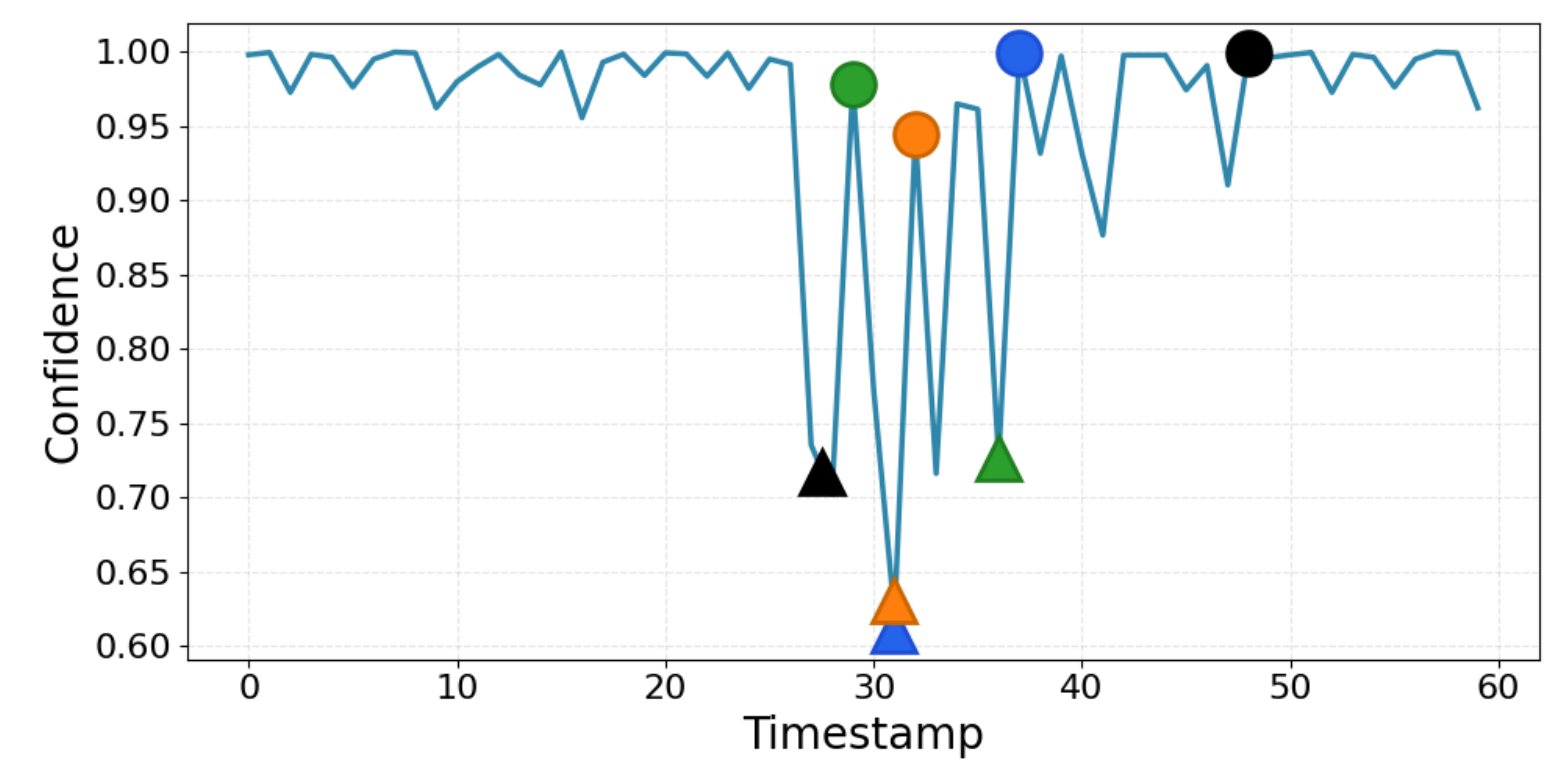}
		\caption{Change of confidence score across model snapshots (SimpleCNN model, CIFAR-10 dataset). The target sample was initially present in the training set. It was removed at the 27th timestamp, and re-inserted at the 48th timestamp.}
        \label{figure:cifar10_conf_timeseries}

\end{figure}

Figure~\ref{figure:cifar10_conf_timeseries} shows a target sample's confidence across 60 SimpleCNN snapshots under the CIFAR-10 update setting in Section~\ref{sc:characterization}. Confidence fluctuates after deletion but remains high and stable after insertion. All three existing methods fail to identify the true entry and exit points, consistent with Section~\ref{sec-motivation}.

\section{Accuracy of Inferred Entry \& Exit Points}
\label{appendix:inference-entry-exit}
\begin{figure*}[!t]
    \centering
     \includegraphics[width=0.45\textwidth,trim=0 20bp 0 20bp,clip]{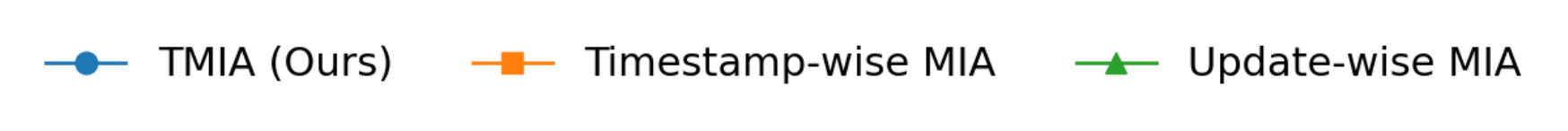}

     \textbf{Entry points}
     \\
     
    \subfloat[SST5/Pythia-70M]{%
    \begin{minipage}[b]{0.24\textwidth}
        \includegraphics[width=0.90\linewidth,trim=0 8bp 0 8bp,clip]{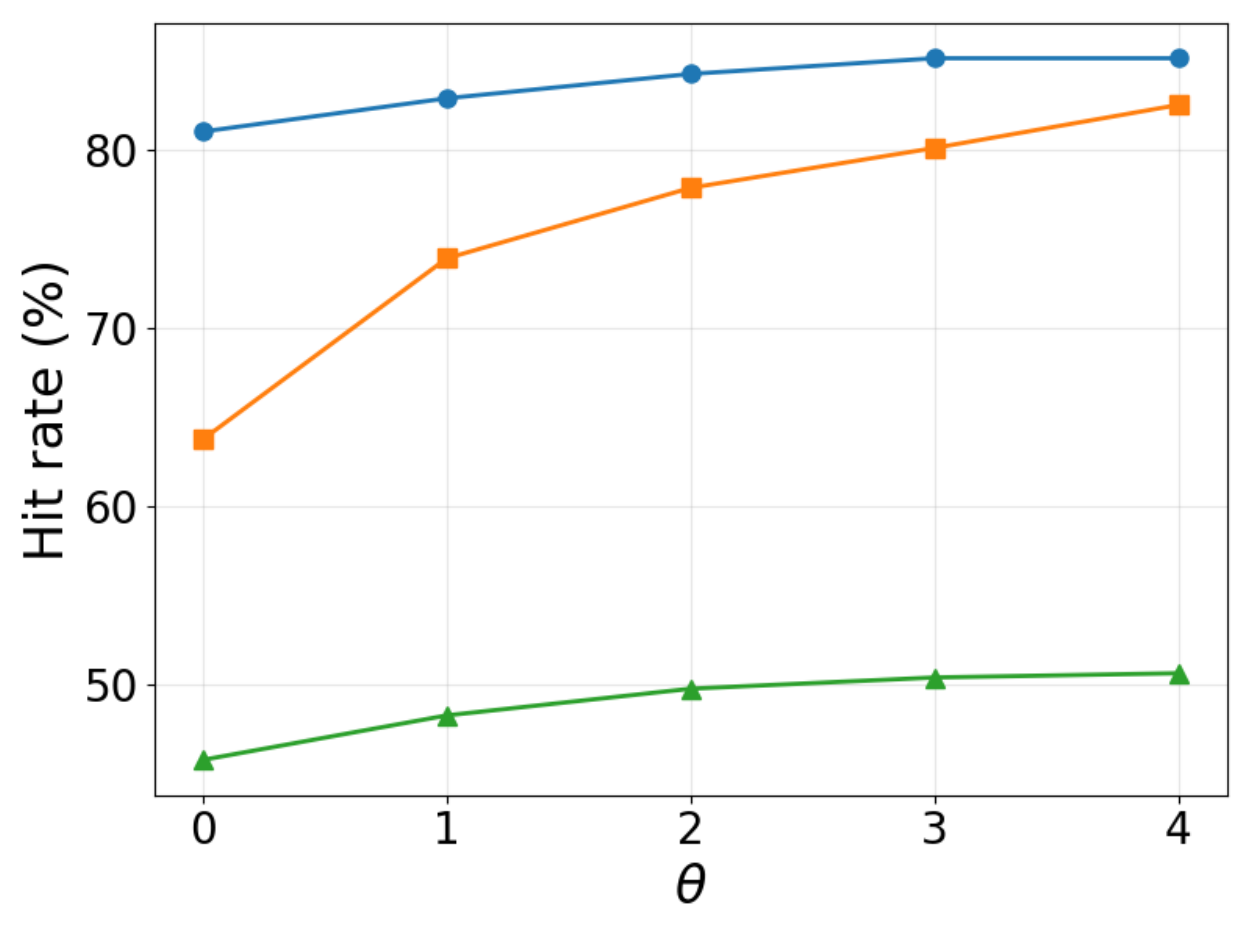}

\label{fig:break_acc_sst5_pyth}
    \end{minipage}%
    }
     \subfloat[News20/Pythia-70M]{%
     \begin{minipage}[b]{0.24\textwidth}
        \includegraphics[width=0.90\linewidth,trim=0 8bp 0 8bp,clip]{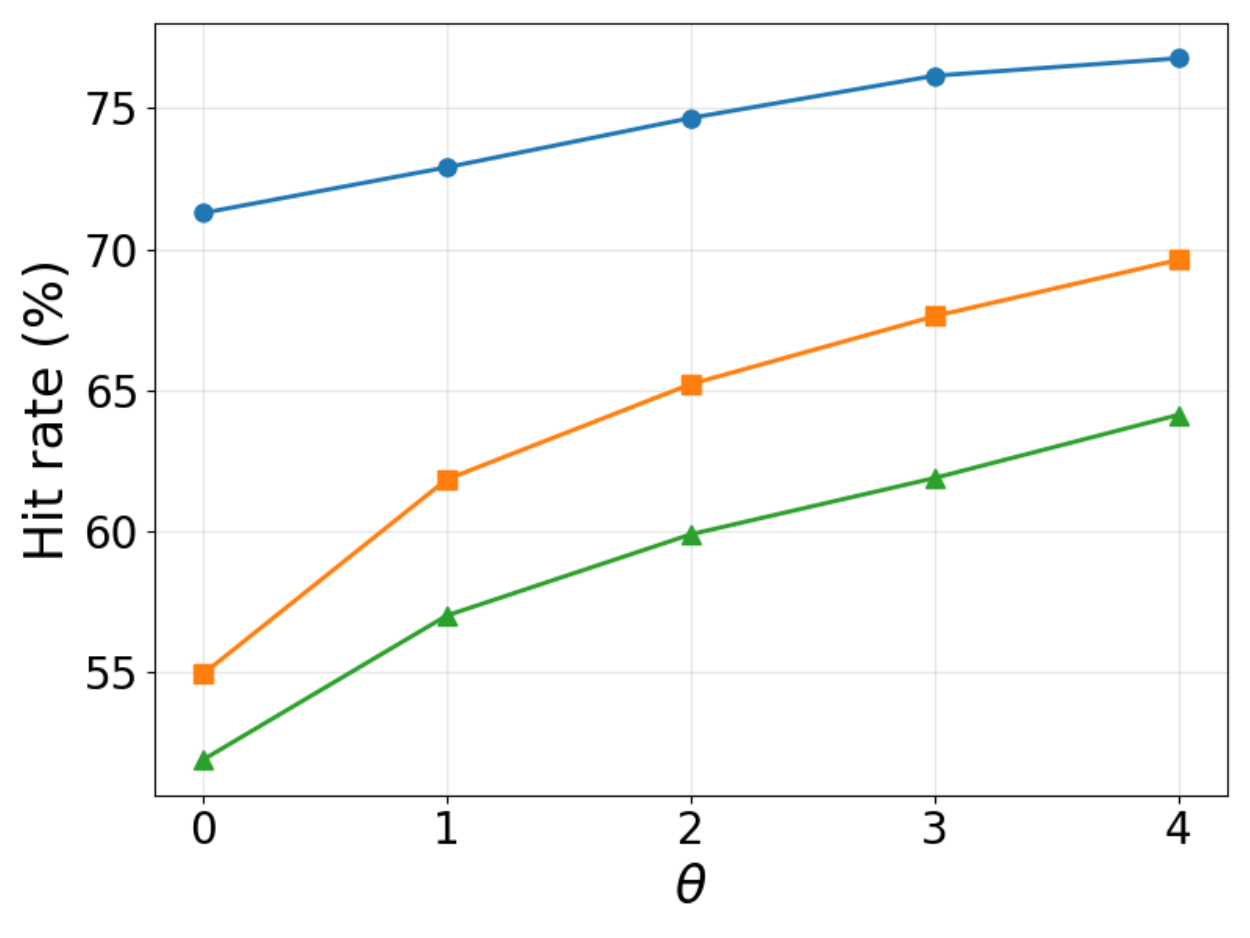}

\label{fig:break_acc_news20_pyth}
     \end{minipage}%
     }
           \subfloat[CIFAR10/Resnet-18]{%
           \begin{minipage}[b]{0.24\textwidth}
        \includegraphics[width=0.90\linewidth,trim=0 8bp 0 8bp,clip]{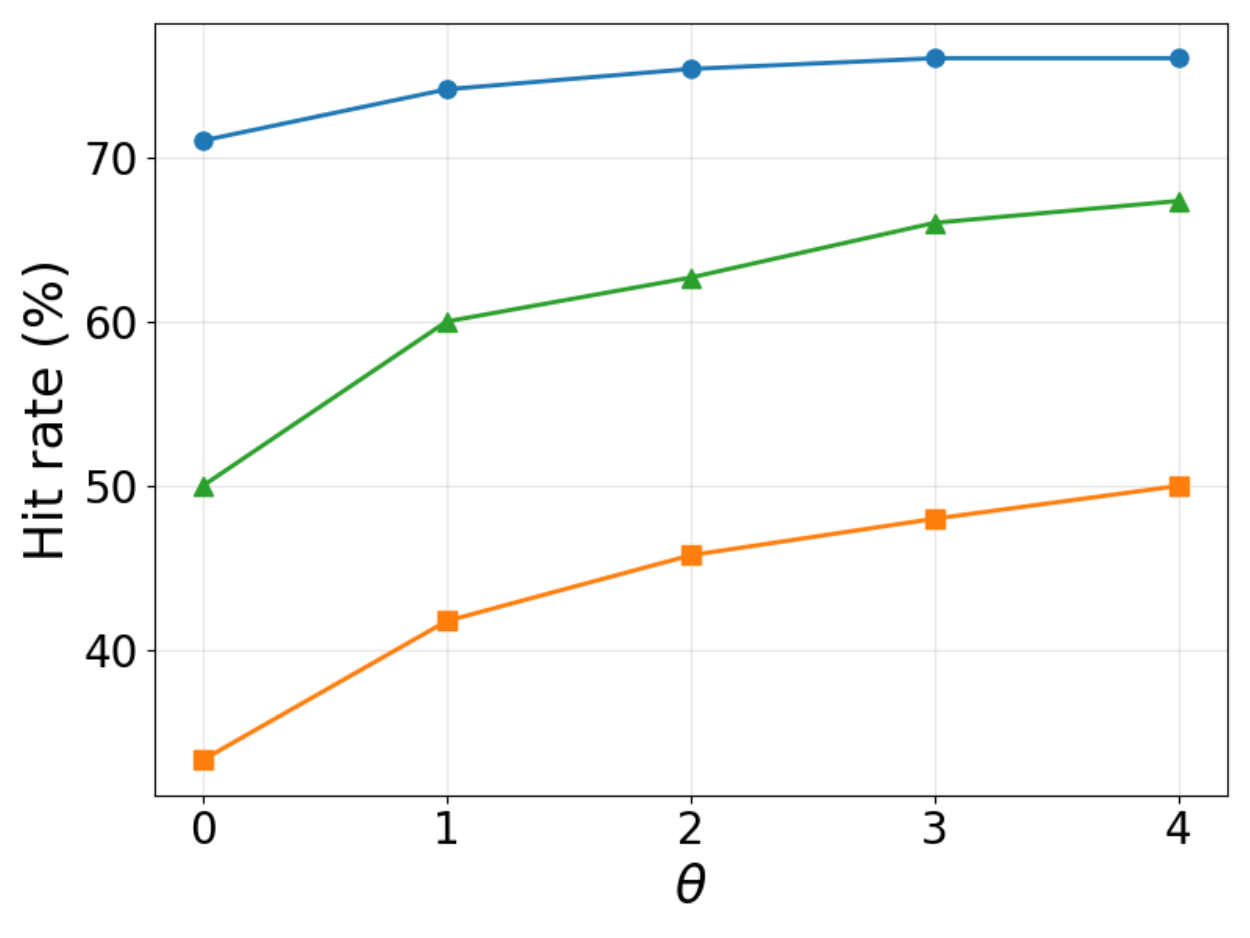}

\label{fig:break_acc_cifar10_resnet}
           \end{minipage}%
           }
         \subfloat[CINIC10/Resnet-18]{%
         \begin{minipage}[b]{0.24\textwidth}
        \includegraphics[width=0.90\linewidth,trim=0 8bp 0 8bp,clip]{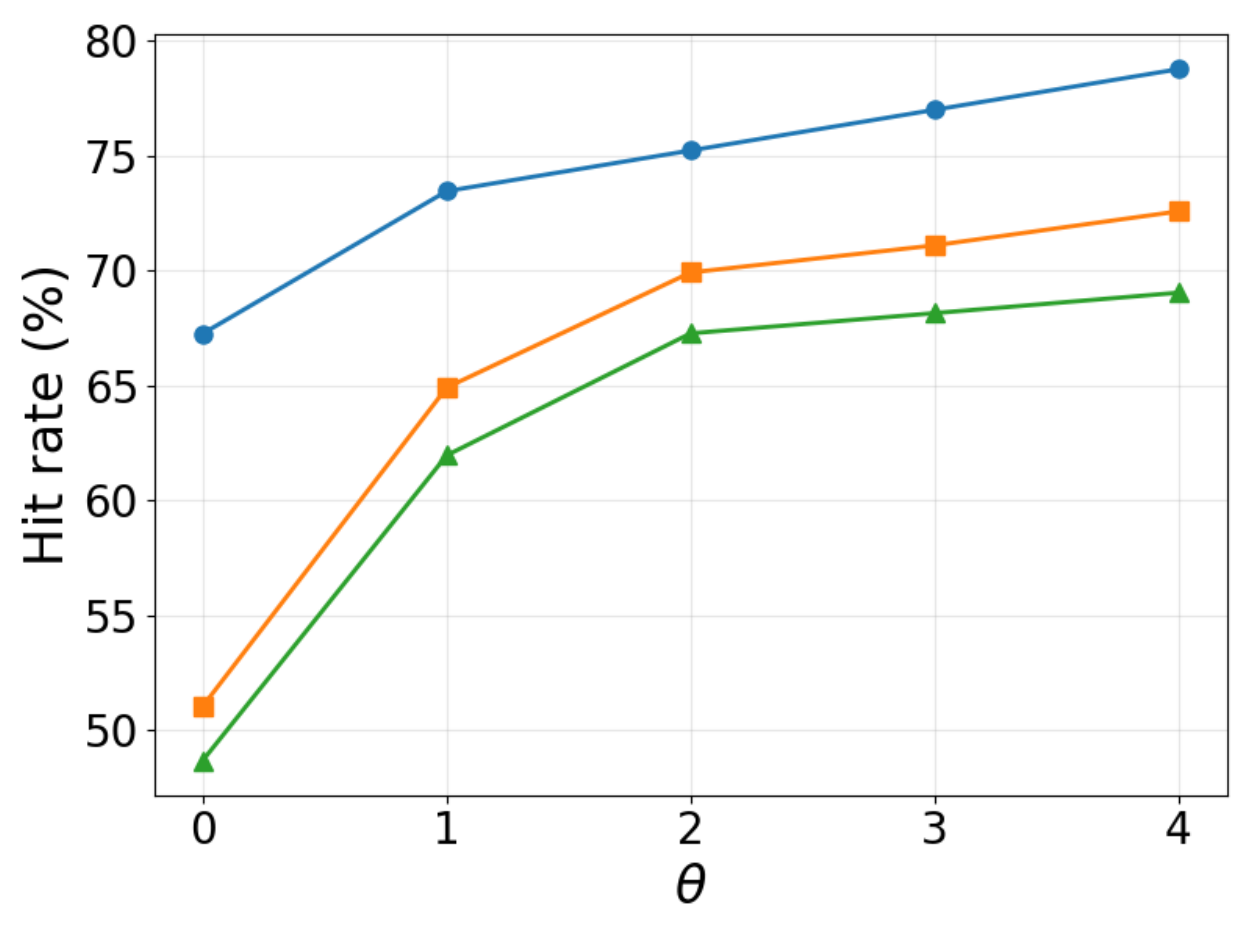}

\label{fig:break_acc_cinic_resnet}
         \end{minipage}%
         }

     \textbf{Exit points}
     \\
     
        \subfloat[SST5/Pythia-70M]{%
        \begin{minipage}[b]{0.24\textwidth}
        \includegraphics[width=0.90\linewidth,trim=0 8bp 0 8bp,clip]{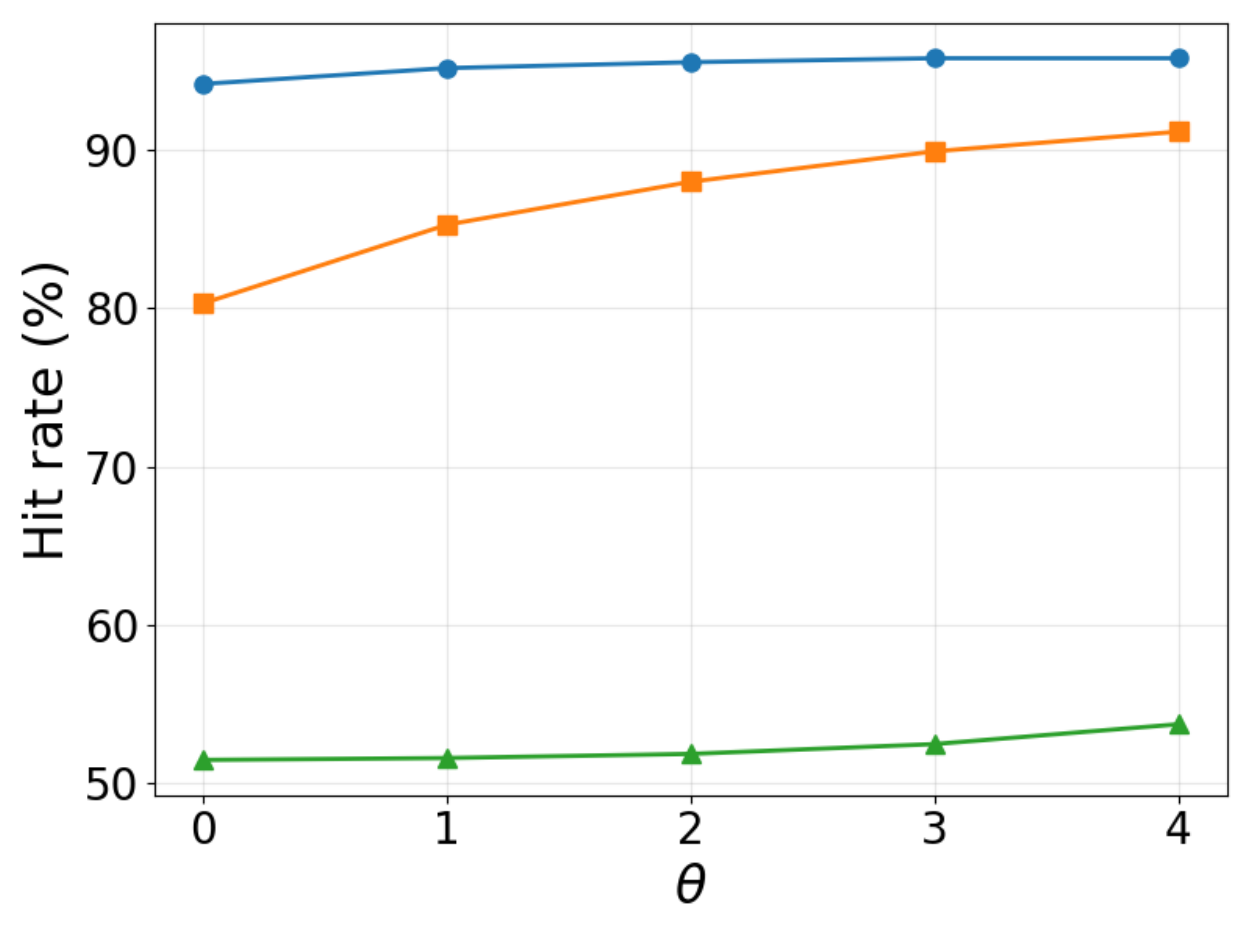}

\label{fig:break_remove_acc_sst5_pyth}
        \end{minipage}%
        }
     \subfloat[News20/Pythia-70M]{%
     \begin{minipage}[b]{0.24\textwidth}
        \includegraphics[width=0.90\linewidth,trim=0 8bp 0 8bp,clip]{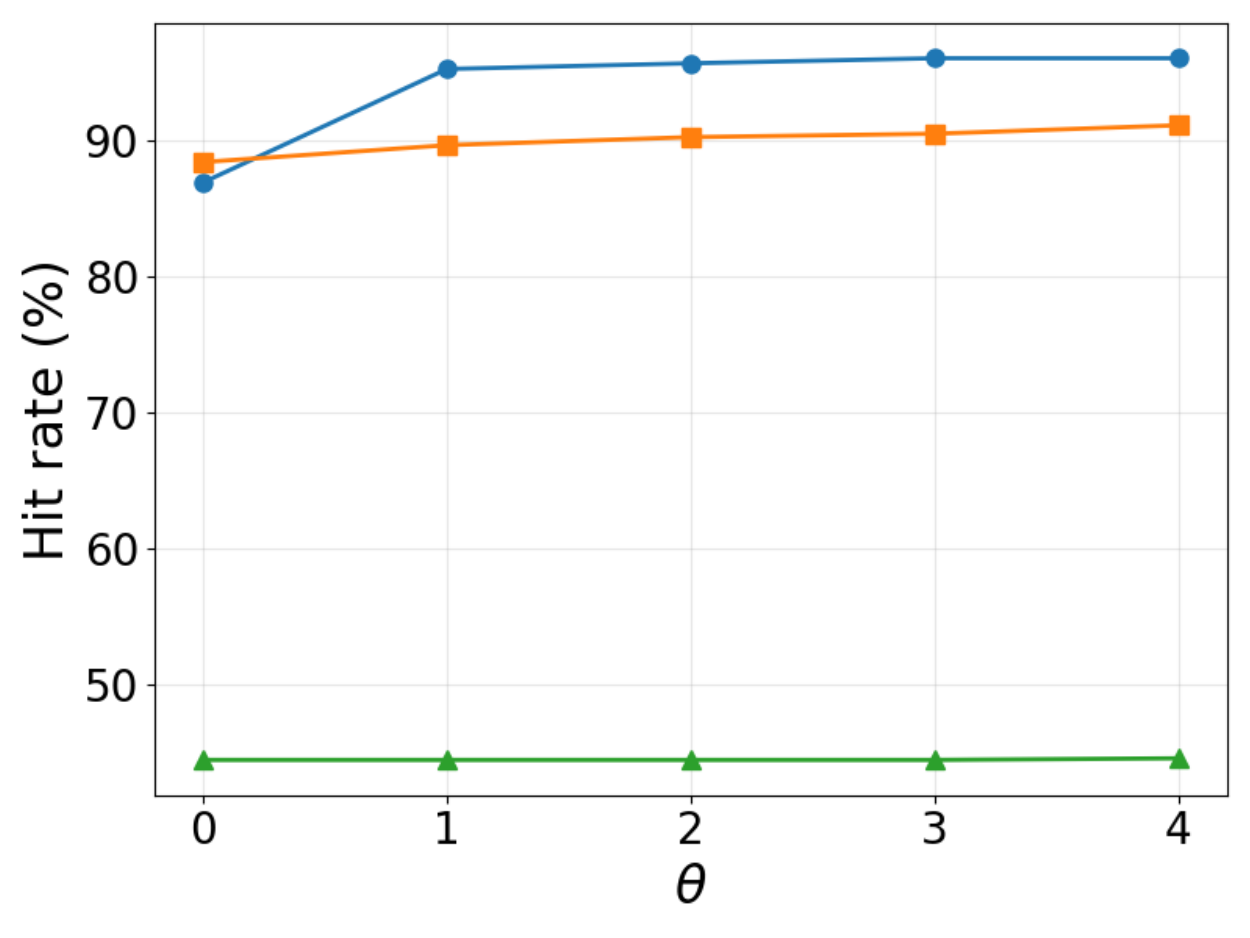}

\label{fig:break_remove_acc_news20_pyth}
     \end{minipage}%
     }
           \subfloat[CIFAR10/Resnet-18]{%
           \begin{minipage}[b]{0.24\textwidth}
        \includegraphics[width=0.90\linewidth,trim=0 8bp 0 8bp,clip]{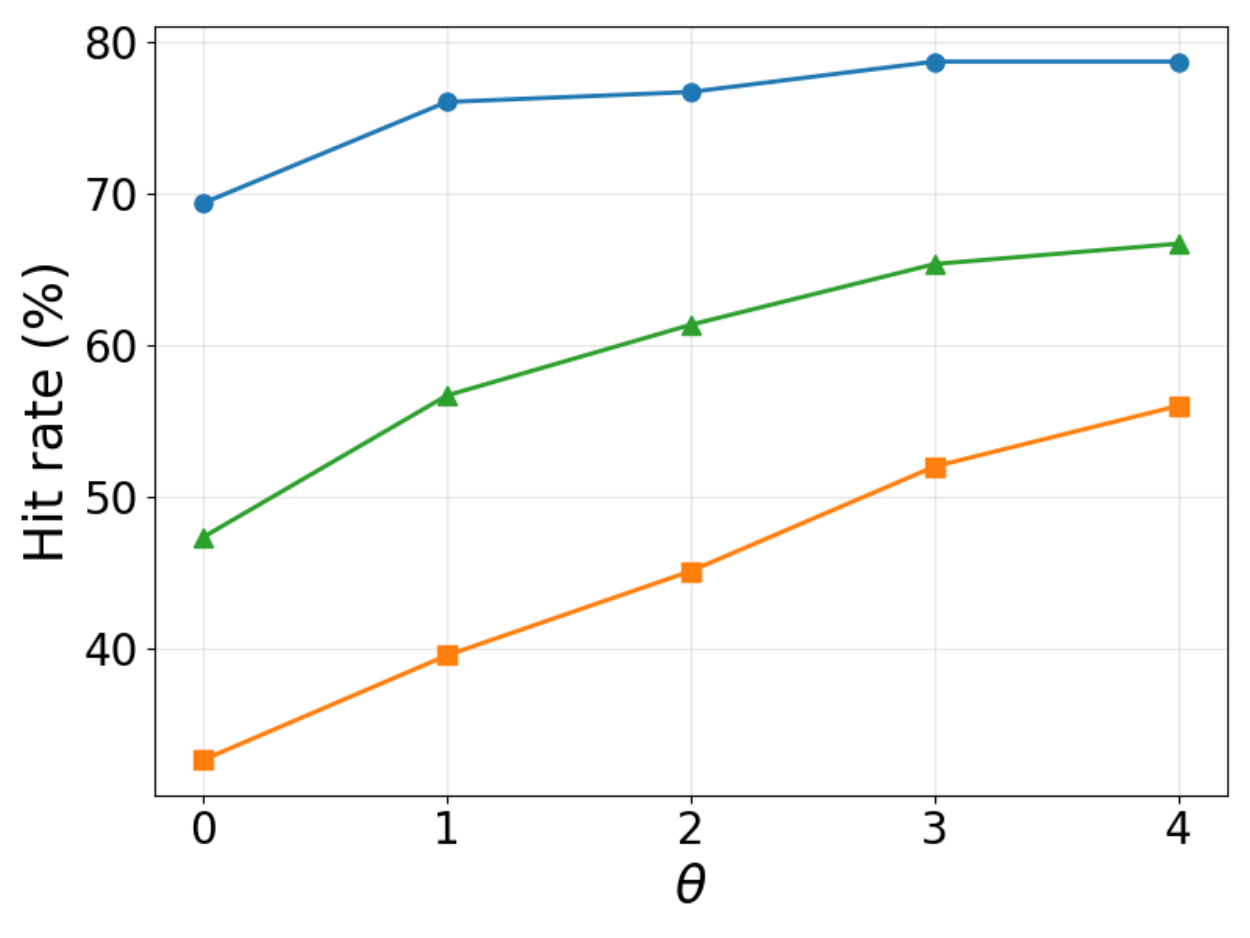}

\label{fig:break_remove_acc_cifar10_resnet}
           \end{minipage}%
           }
         \subfloat[CINIC10/Resnet-18]{%
         \begin{minipage}[b]{0.24\textwidth}
        \includegraphics[width=0.90\linewidth,trim=0 8bp 0 8bp,clip]{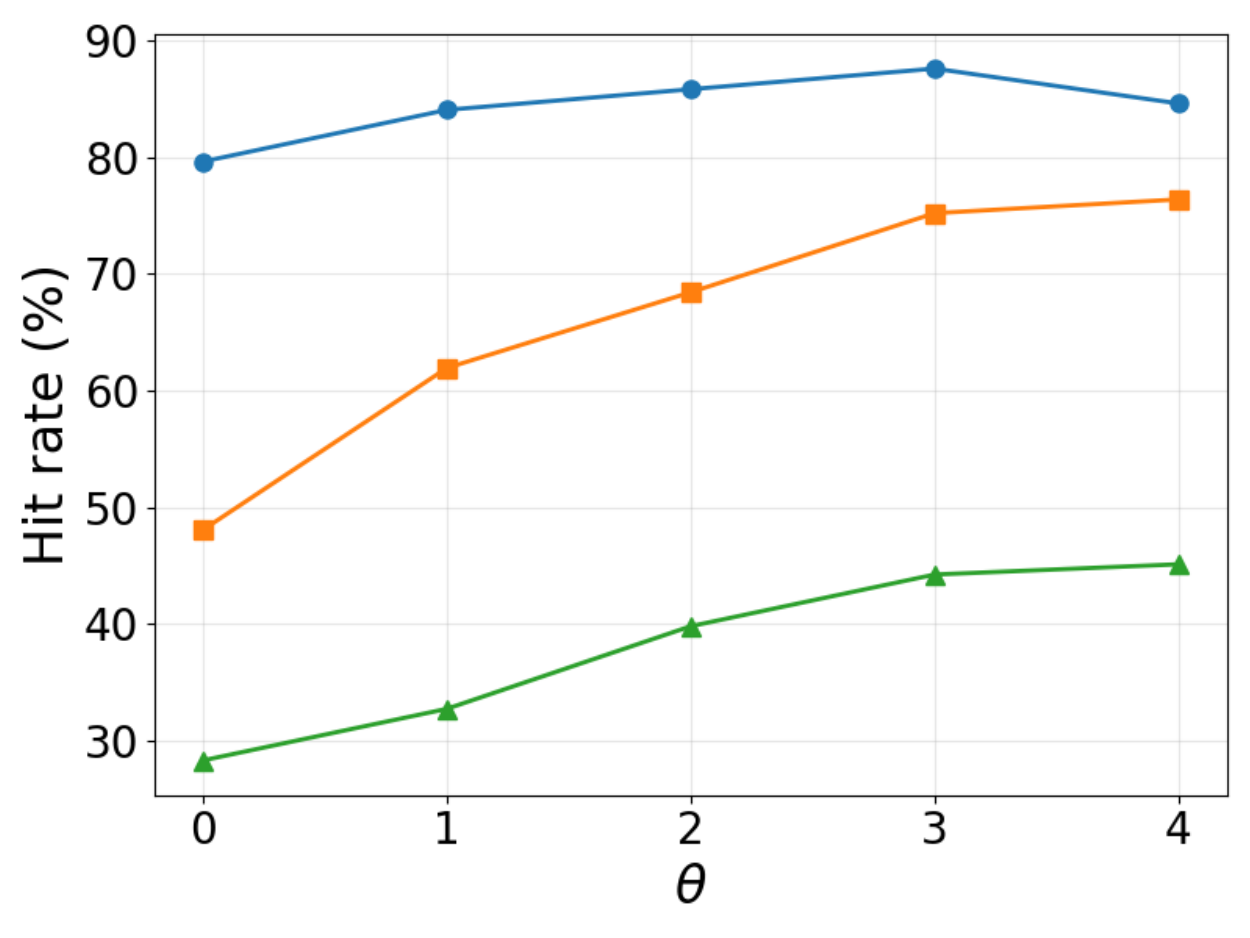}

\label{fig:break_remove_acc_cinic_resnet}
         \end{minipage}%
         }
    
     \caption{Accuracy of inferring entry and exit points} 
    \label{figure:Breakpoints-detection-Accuracy}
\end{figure*}

Evaluating the accuracy of inferred \emph{entry} and \emph{exit} points is not straightforward because these inferred points may not exactly match their ground-truth ones but with a small derivation instead. 
Consequently, even when the predicted entry or exit points does not exactly coincide with their ground-truth, a prediction with a small temporal offset should still be regarded as successfully localizing the membership transition in time. Therefore, we introduce a \textit{tolerance window} to enable a more realistic evaluation of attack effectiveness in continually updated settings, without overly penalizing minor temporal misalignments. 

Specifically, give a tolerance window of size $\theta$, we say a ground-truth entry/exit point $t^*_j$ has a \textit{hit} if there exists a predicted entry/exit point $t_i$ within the tolerance window, i.e., $t_i\in[t^*_j-\theta, t^*_j+\theta]$.  
Based on the definition of the hit,  we measure the {\em hit rate} of the ground-truth entry/exit points as the percentage of these points that are correctly predicted. 
We report the hit rate of entry and exit points by \system\ and two baseline methods in 
Figure~\ref{figure:Breakpoints-detection-Accuracy}. 

\textbf{Comparison with baselines.}
Our method consistently outperforms both baseline approaches across all settings.
The performance gap is particularly pronounced at small $\theta$, where
accurate temporal localization is most critical.
For instance, at $\theta = 0$, \system\ substantially improves the hit rate for entry points compared to the timestamp-wise MIA, achieving 71.00\% versus 33.33\% when attacking the ResNet-18 model trained on CIFAR-10 dataset, and 81.04\% versus 63.75\% when attacking the Pythia-70M trained on SST-5 dataset. Similar gains are observed for exit point detection. As $\theta$ increases, \system\ consistently achieves higher hit rates than the baseline methods, 
indicating that it not only detects correct entry and exit events more
frequently, but also localizes them more precisely in time.

\textbf{Effect of tolerance window size.}
Across all datasets and models, increasing the tolerance window $\theta$
consistently improves the hit rate for methods that produce temporally
localized entry and exit predictions, including \system\ and
Timestamp-wise MIA.
As $\theta$ increases from 0 to 4, both methods exhibit a monotonic or
near-monotonic improvement in hit rate for both entry and exit detection.
This trend indicates that many prediction errors correspond to small
temporal shifts, rather than completely incorrect inferences, and that a
larger tolerance window can absorb minor misalignments between predicted
and ground-truth transition points. In contrast, while Update-wise MIA also shows modest improvements in hit
rate as $\theta$ increases in certain settings (e.g., CIFAR-10 and
CINIC-10), its gains are substantially smaller and less consistent.
This suggests that Update-wise MIA lacks precise temporal localization,
and its predictions remain weakly aligned with the true entry and exit
points even under relaxed tolerance.

\section{Additional Results: Robustness of \system\  Under Shadow Model Heterogeneity}
\label{appendix:shadow-model-mismatch}

\subsection{Interval-level Attack Performance}
\label{appendix-I-F1-results}

\begin{figure*}[!t]
    \centering
         \textbf{\small I-F1}
     \\
            \subfloat[Target-shadow datasets]{%
            \begin{minipage}[b]{0.245\textwidth}
        \includegraphics[width=\textwidth]{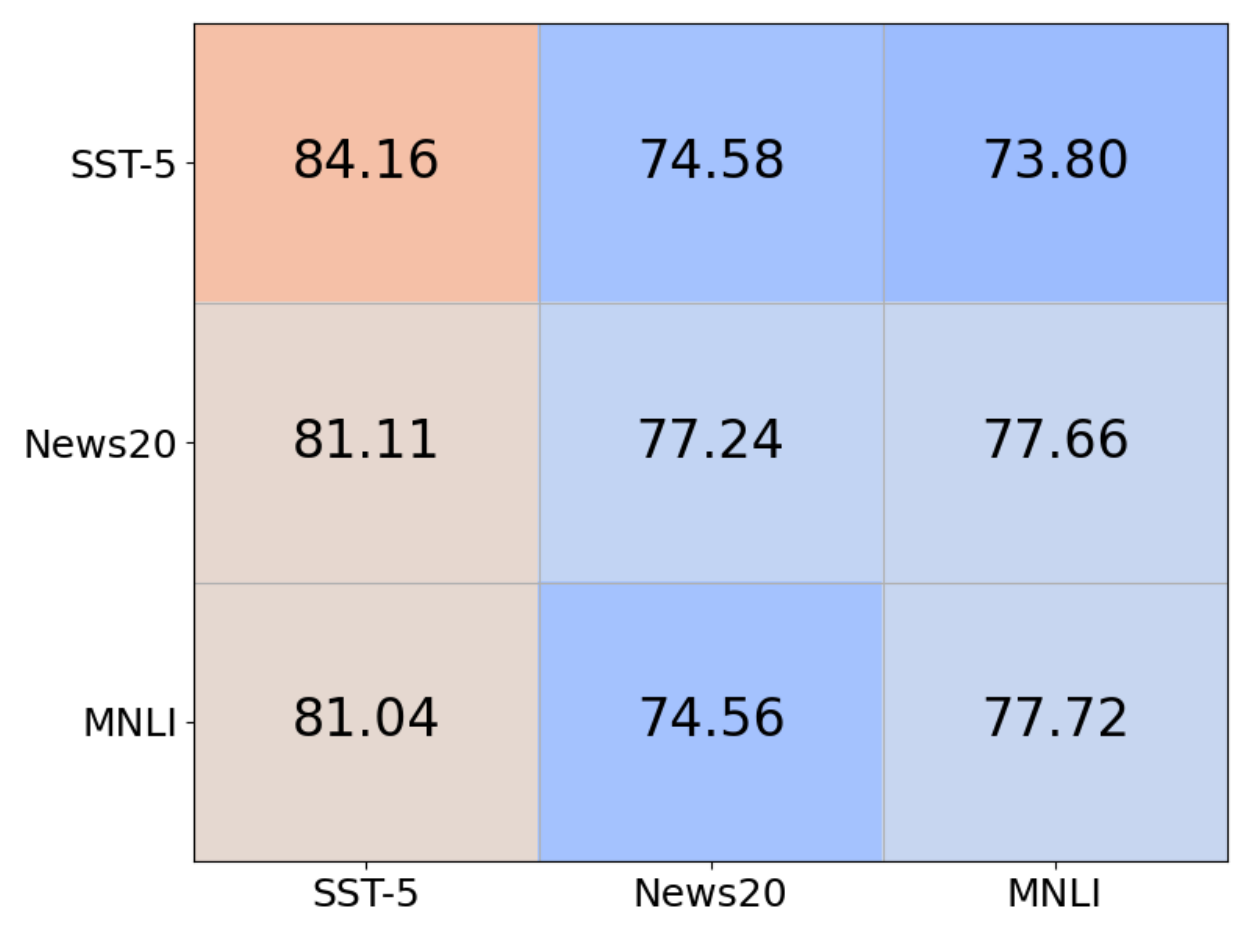}
        \label{fig:plot_dataset_transfer_f1_interval}
            \end{minipage}%
            }
    \subfloat[Target-shadow models]{%
    \begin{minipage}[b]{0.245\textwidth}
        \includegraphics[width=\textwidth]{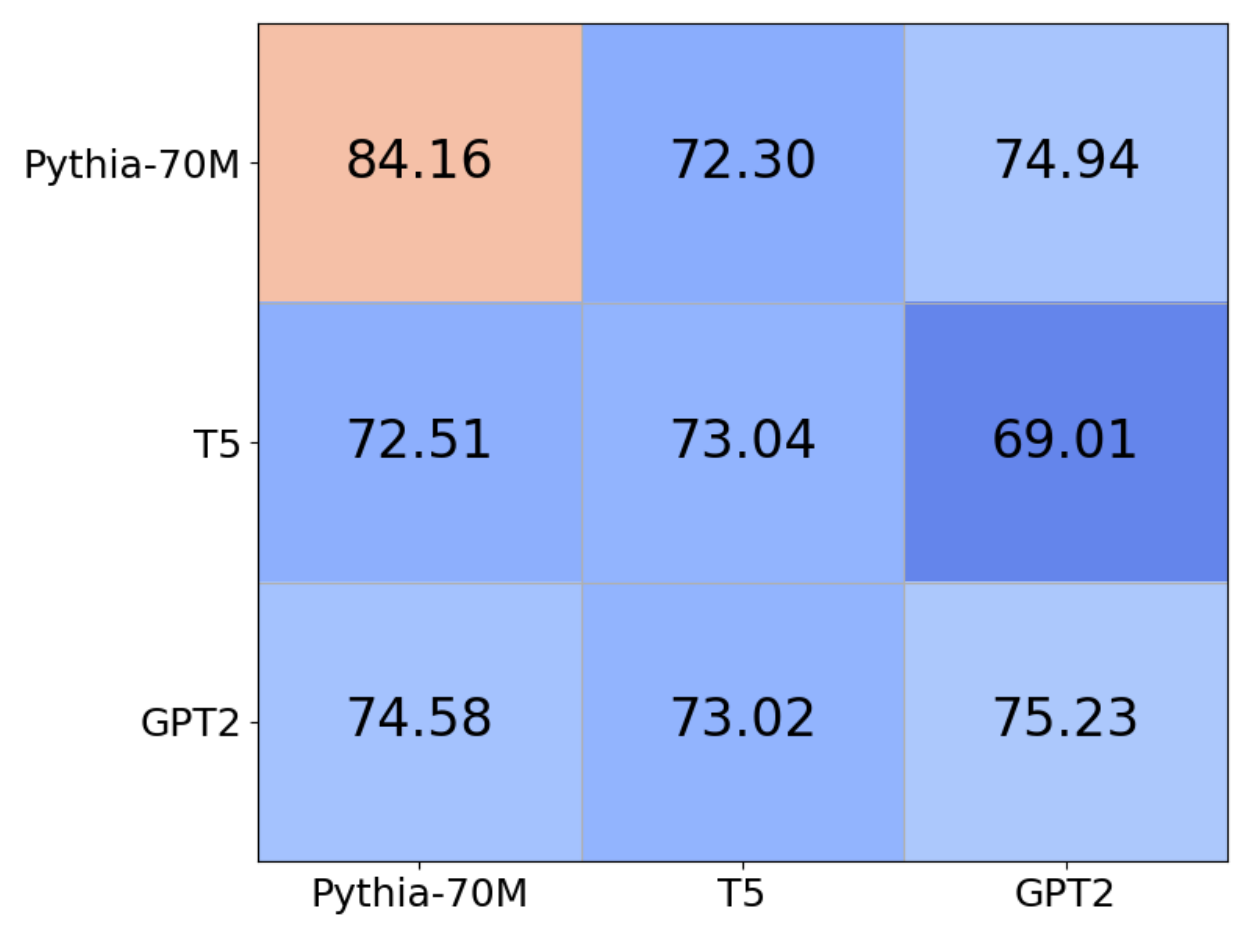} 
\label{fig:plot_model_transfer_f1_interval}
    \end{minipage}%
    }
     \subfloat[Update methods]{%
     \begin{minipage}[b]{0.245\textwidth}
        \includegraphics[width=\textwidth]{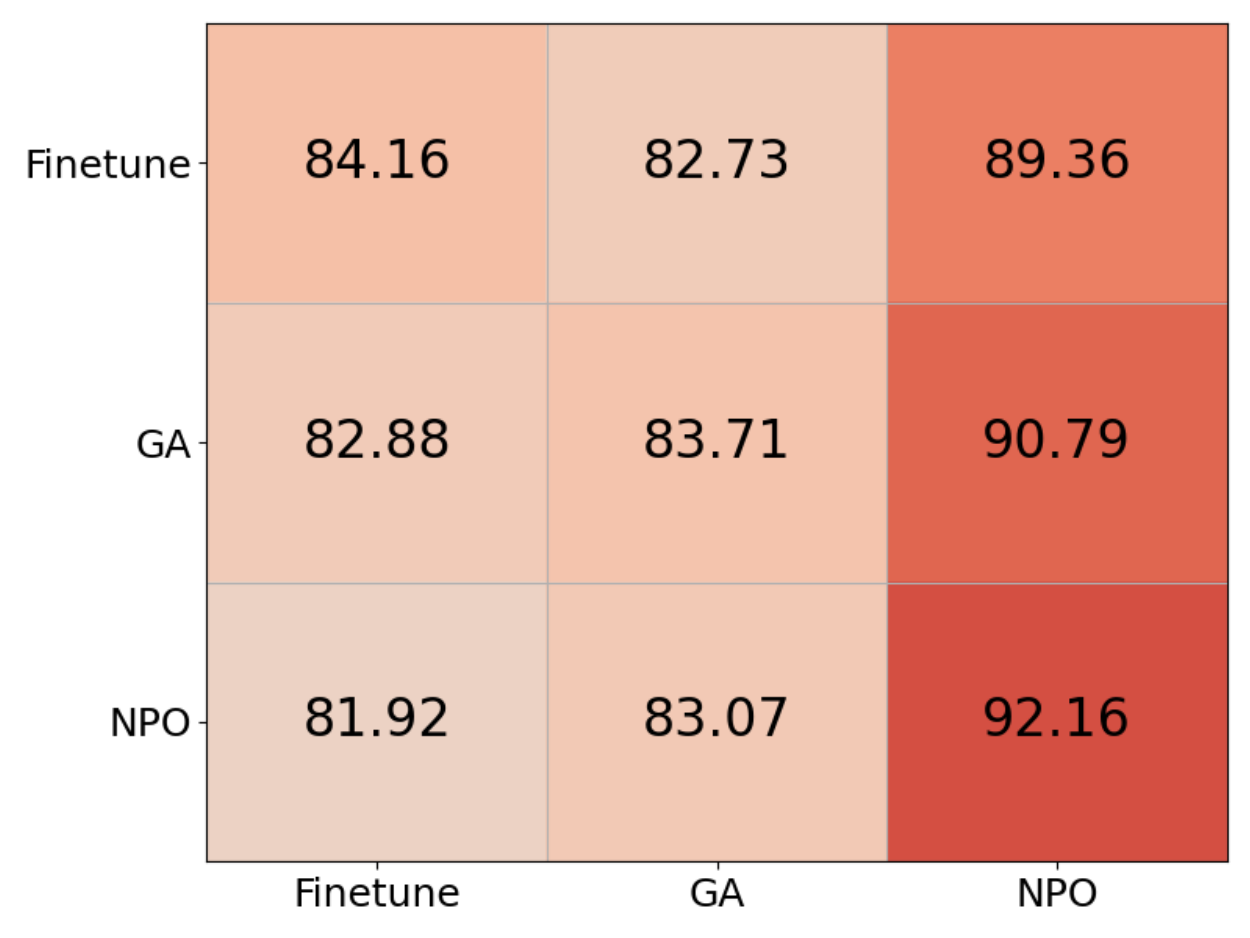}
\label{fig:plot_unlearning_transfer_f1_interval}
     \end{minipage}%
     }
    \subfloat[Update volume]{%
     \begin{minipage}[b]{0.245\textwidth}
        \includegraphics[width=\textwidth]{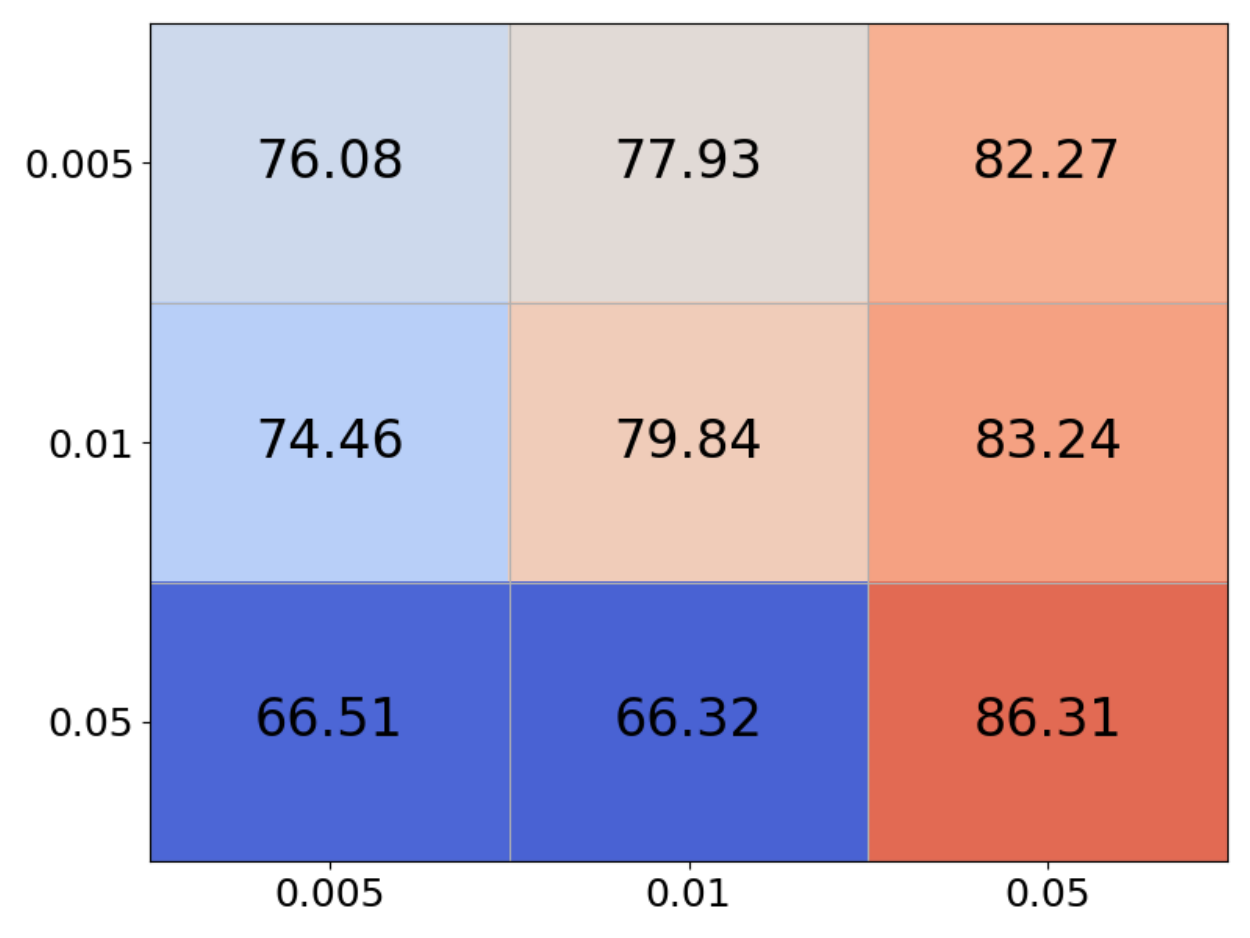}
        \label{fig:Target-shadow update size_I_f1}
     \end{minipage}}
   \caption{Interval-level attack Performance of \system\ when the shadow model mismatches the target model. The horizontal and vertical axes correspond to the target and shadow configurations, respectively, including the model, dataset, update method, and update volume.}
  \label{figure:Attack-transferability-I-F1}
\end{figure*}

Figure~\ref{figure:Attack-transferability-I-F1} reports interval-level performance under shadow-model mismatch for the three LMs.

\noindent\textbf{Mismatched Datasets.}
With Pythia-70M fixed, we vary shadow and target datasets among SST-5, News20, and MNLI. Matching datasets yield the highest I-F1; mismatch reduces I-F1 by at most 5.04\% (Figure~\ref{fig:plot_dataset_transfer_f1_interval}).

\noindent\textbf{Mismatched Model Architectures.}
On SST-5, we vary shadow and target architectures among Pythia-70M, T5, and GPT-2. Matching architectures perform best; the maximum I-F1 reduction under mismatch is approximately 13\% (Figure~\ref{fig:plot_model_transfer_f1_interval}).

\noindent\textbf{Mismatched Update Algorithms.}
For Pythia-70M/SST-5, we vary shadow and target update algorithms. Matching algorithms yield the highest I-F1, while mismatched configurations retain I-F1 of at least 81.92\% (Figure~\ref{fig:plot_unlearning_transfer_f1_interval}).

\noindent\textbf{Mismatched Update Volume.}
On SST-5/Pythia-70M, we vary shadow and target per-timestamp update volumes among $0.5\%$, $1\%$, and $5\%$ of the training-set size (Figure~\ref{fig:Target-shadow update size_I_f1}). For each target, matching volumes maximize I-F1. Relative to matched settings, I-F1 drops by at most 4.04 percentage points when the shadow volume does not exceed the target's. 

\begin{figure*}[!t]
    \centering
     \includegraphics[width=0.26\textwidth]{figures/parameter_legend.pdf}
     
    \subfloat[\small Sliding-window size]{%
    \begin{minipage}[b]{0.24\textwidth}
        \includegraphics[width=0.90\linewidth,trim=0 8bp 0 8bp,clip]{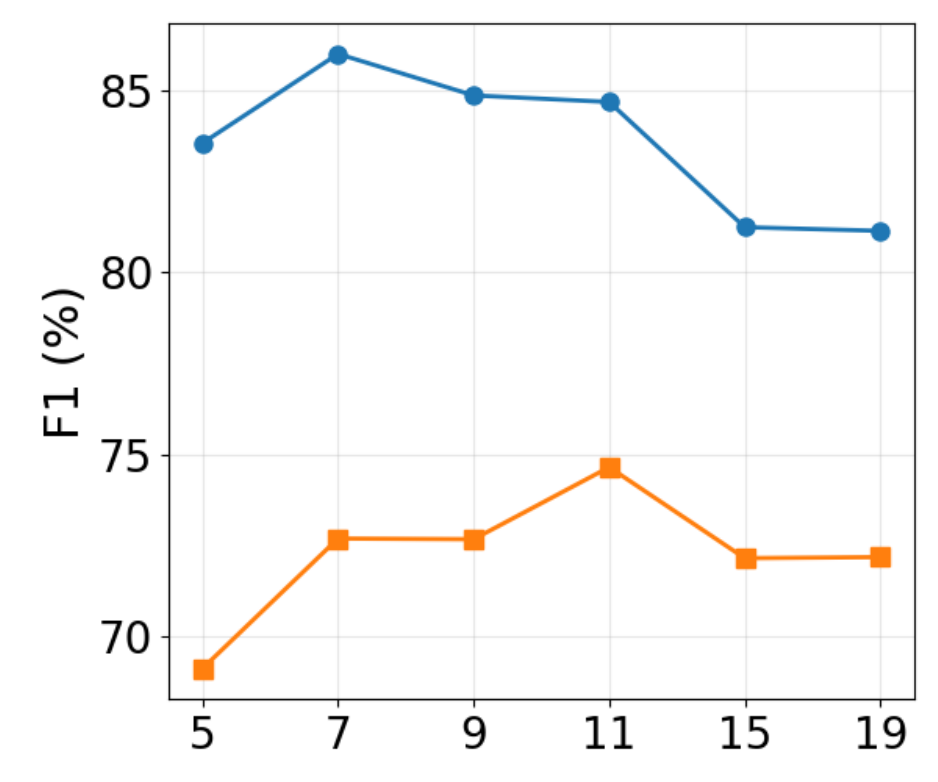}

\label{fig:parameter_window_size_news20}
    \end{minipage}%
    }
     \subfloat[\small Update volume]{%
     \begin{minipage}[b]{0.24\textwidth}
        \includegraphics[width=0.90\linewidth,trim=0 8bp 0 8bp,clip]{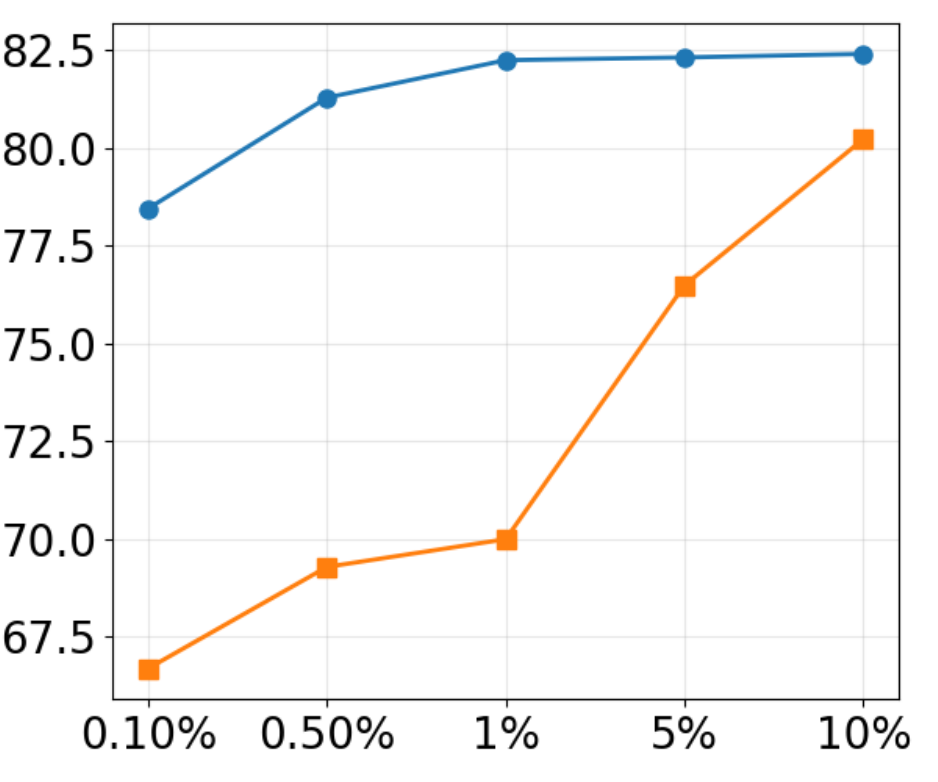}

\label{fig:parameter_update_size_news20}
     \end{minipage}%
     }
           \subfloat[\small Update count]{%
           \begin{minipage}[b]{0.24\textwidth}
        \includegraphics[width=0.90\linewidth,trim=0 8bp 0 8bp,clip]{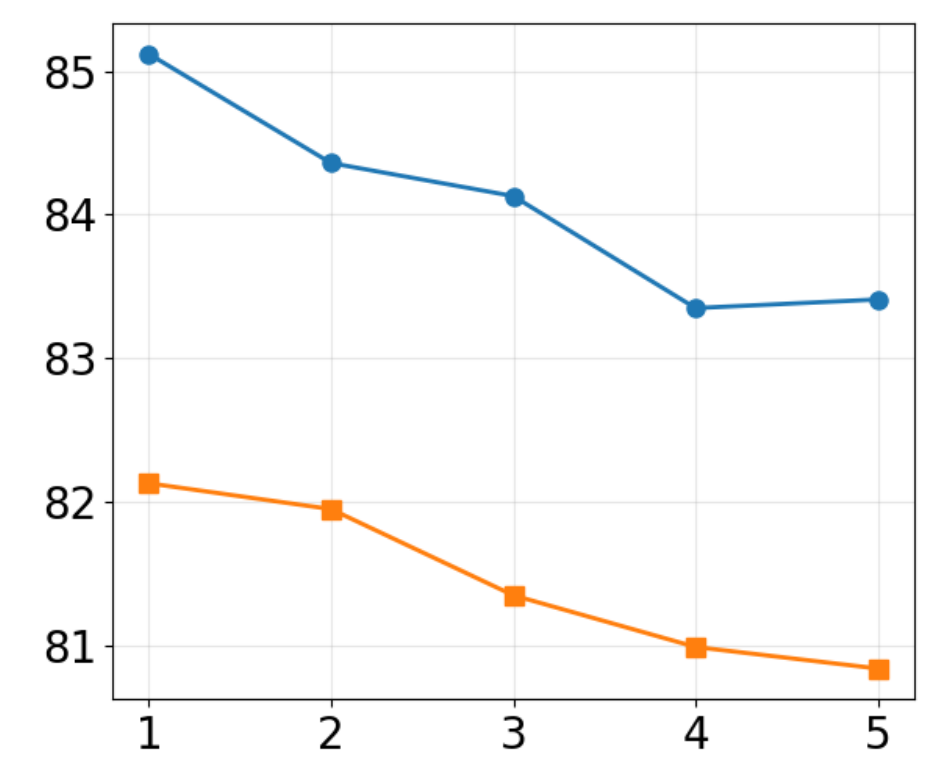}

\label{fig:parameter_insert_news20}
           \end{minipage}%
           }
         \subfloat[\small Membership duration]{%
         \begin{minipage}[b]{0.24\textwidth}
        \includegraphics[width=0.90\linewidth,trim=0 8bp 0 8bp,clip]{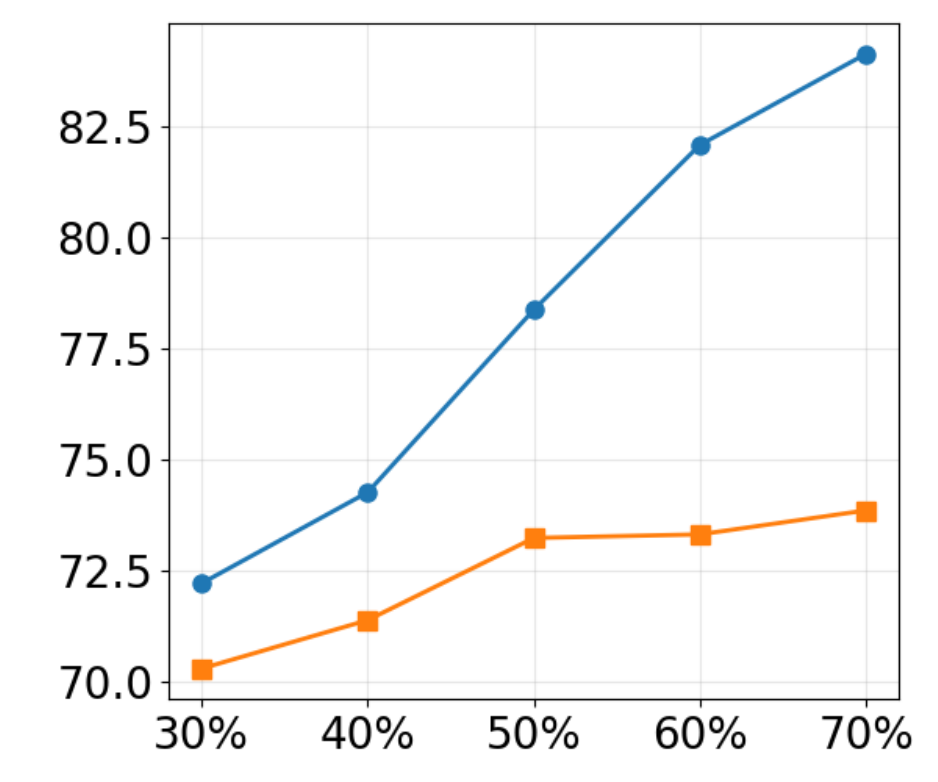}

\label{fig:parameter_interval_member_news20}
         \end{minipage}%
         }
     \caption{Performance of \system\ under various parameter configurations (Pythia-70M, News20)}  
    \label{figure:factor-analysis-news20}
\end{figure*}

\begin{table*}[!t]
\centering
\small
\caption{Interval-level acc of \system\ and baselines on generative models. The best results for each setting are marked in \gray{gray}. }
\begin{tabular}{c|c|ccccccccc}
\hline
\multirow{2}{*}{\textbf{Model}} & \multirow{2}{*}{\textbf{Dataset}} & \multicolumn{3}{c|}{\textbf{TMIA (Ours)}} 
                                                                    & \multicolumn{3}{c|}{\textbf{Timestamp-wise MIA}} 
                                                                                                            & \multicolumn{3}{c}{\textbf{Update-wise MIA}} \\ \cline{3-11} 
                                &                                   & \multicolumn{1}{c|}{\textbf{I-Pre}} 
                                                                    & \multicolumn{1}{c|}{\textbf{I-Rec}} 
                                                                                                            & \multicolumn{1}{c|}{\textbf{I-F1}} 
                                                                                                                                    & \multicolumn{1}{c|}{\textbf{I-Pre}} 
                                                                                                                                                                            & \multicolumn{1}{c|}{\textbf{I-Rec}} 
                                                                                                                                                                                                                    & \multicolumn{1}{c|}{\textbf{I-F1}} 
                                                                                                                                                                                                                                            & \multicolumn{1}{c|}{\textbf{I-Pre}} 
                                                                                                                                                                                                                                                                    & \multicolumn{1}{c|}{\textbf{I-Rec}} 
                                                                                                                                                                                                                                                                                                            & \textbf{I-F1} \\ \hline
\multirow{2}{*}{Llama-3B}       
& Squad     
& \multicolumn{1}{c|}{97.37} 
& \multicolumn{1}{c|}{82.59} 
& \multicolumn{1}{c|}{\gray{89.37}} 
& \multicolumn{1}{c|}{38.42} 
& \multicolumn{1}{c|}{84.67} 
& \multicolumn{1}{c|}{52.86} 
& \multicolumn{1}{c|}{34.30} 
& \multicolumn{1}{c|}{10.90} 
& 16.54 \\
                                
& Simpleqa  
& \multicolumn{1}{c|}{97.23} 
& \multicolumn{1}{c|}{92.89} 
& \multicolumn{1}{c|}{\gray{95.01}} 
& \multicolumn{1}{c|}{34.98} 
& \multicolumn{1}{c|}{59.15} 
& \multicolumn{1}{c|}{43.96} 
& \multicolumn{1}{c|}{60.01} 
& \multicolumn{1}{c|}{8.84} 
& 15.41 \\ \hline
\multirow{2}{*}{Llama-8B}       
& Squad     
& \multicolumn{1}{c|}{94.63} 
& \multicolumn{1}{c|}{87.14} 
& \multicolumn{1}{c|}{\gray{90.73}} 
& \multicolumn{1}{c|}{35.32} 
& \multicolumn{1}{c|}{82.68} 
& \multicolumn{1}{c|}{49.50} 
& \multicolumn{1}{c|}{33.86} 
& \multicolumn{1}{c|}{10.55} 
& 16.09 \\
                                
& Simpleqa  
& \multicolumn{1}{c|}{96.46} 
& \multicolumn{1}{c|}{73.22} 
& \multicolumn{1}{c|}{\gray{83.25}} 
& \multicolumn{1}{c|}{31.71} 
& \multicolumn{1}{c|}{49.91} 
& \multicolumn{1}{c|}{38.78} 
& \multicolumn{1}{c|}{58.82} 
& \multicolumn{1}{c|}{8.35} 
& 14.62 \\ \hline
\end{tabular}
\label{tab:LLM-IL}  
\end{table*}

\section{Additional Results: Parameter Sensitivity Analysis} \label{Appendix:Parameter-Sensitivity-Analysis}

Figure~\ref{figure:factor-analysis-news20} reports \system's performance under various parameter configurations, with Pythia-70M trained on News20 dataset as the target model. 

\textbf{Sliding-window Size}
Figure~\ref{fig:parameter_window_size_news20} presents the impact of different sliding-window sizes on the attack performance of \system\ on the News20 dataset. The overall trend remains consistent with our previous experiments: as the sliding-window size increases, the attack performance improves steadily. Specifically, a window size of 7 yields the best T-F1 performance, while a window size of 11 achieves the highest I-F1 score. 

\textbf{Per-timestamp Update Volume}
We further examine the effect of different update volumes, measured as a ratio of the training data size ranging from 0.1\% to 10\%. As shown in Figure~\ref{fig:parameter_update_size_news20}, both T-F1 and I-F1 scores increase as the per-timestamp update volume grows. As a result, samples form more stable and contiguous temporal intervals
in their member and non-member states, amplifying the temporal signals
exploited by \system.

\textbf{Per-sample Update Count}
Figure~\ref{fig:parameter_insert_news20} reports the performance of \system\ under varying per-sample update counts, where the number of insertion--removal cycles increases from 1 to 5. We observe a gradual decrease in both T-F1 and I-F1 as the update count increases. 

\textbf{Per-sample Membership Duration}
Figure~\ref{fig:parameter_interval_member_news20} illustrates \system’s performance under different membership durations, measured as the ratio of the membership interval to the total number of timestamps (80 in total). The results show that \system’s performance consistently improves as the membership duration increases across all settings. 

\section{Additional Results: Interval-level attack performance for Large Generative Models}\label{appendix:I-F1-LLM}
Table~\ref{tab:LLM-IL} compares interval-level attacks on large generative models. \system\ achieves strong I-Pre and I-Rec, with I-F1 reaching 95.01\%, while both baselines yield substantially lower I-F1. These results show that \system\ recovers continuous membership intervals even from large generative models.

\bibliographystyle{ACM-Reference-Format}
\bibliography{cite}
\end{document}